\documentclass[a4paper,fleqn]{cas-dc}

\usepackage[authoryear]{natbib}
\usepackage{subcaption}
\usepackage{supertabular} 
\newcommand{\PreserveBackslash}[1]{\let\temp=\\#1\let\\=\temp}
\newcolumntype{C}[1]{>{\PreserveBackslash\centering}p{#1}}
\newcolumntype{R}[1]{>{\PreserveBackslash\raggedleft}p{#1}}
\newcolumntype{L}[1]{>{\PreserveBackslash\raggedright}p{#1}}

\def\tsc#1{\csdef{#1}{\textsc{\lowercase{#1}}\xspace}}
\tsc{WGM}
\tsc{QE}
\tsc{EP}
\tsc{PMS}
\tsc{BEC}
\tsc{DE}

\begin{document}
\let\WriteBookmarks\relax
\def\floatpagepagefraction{1}
\def\textpagefraction{.001}
\shorttitle{Integrating Hydrogen in Single-Price Electricity Systems}
\shortauthors{vom Scheidt et~al.}

\title [mode = title]{Integrating Hydrogen in Single-Price Electricity Systems: The Effects of Spatial Economic Signals}

\author[1]{Frederik vom Scheidt}[]
    \cormark[1]
    \ead{frederik.scheidt@kit.edu}
    \ead[url]{frederik.scheidt@kit.edu}
    \credit{Conceptualization, Methodology, Software, Formal analysis, Investigation, Data Curation, Writing - Original Draft, Writing - Review \& Editing, Visualization, Project administration}
    \address[1]{Karlsruhe Institute of Technology, Karlsruhe, Germany}

\author[2]{ Jingyi Qu}[]
    \credit{Software, Resources, Data Curation, Visualization, Writing - Review \& Editing}
    \address[2]{FZI Research Center for Information Technology, Karlsruhe, Germany}

\author[1]{ Philipp Staudt}[]
    \credit{Conceptualization, Writing - Review \& Editing, Supervision, Funding acquisition}
    
\author[3]{ Dharik S. Mallapragada}[]
    \credit{Conceptualization, Writing - Review \& Editing}
    \address[3]{Massachusetts Institute of Technology, Cambridge, USA}
        
\author[1]{ Christof Weinhardt}[]
    \credit{Resources, Supervision, Funding acquisition}

\cortext[cor1]{Corresponding author}

\begin{abstract}
Hydrogen can contribute substantially to the reduction of carbon emissions in industry and transportation. However, the production of hydrogen through electrolysis creates interdependencies between hydrogen supply chains and electricity systems. Therefore, as governments worldwide are planning considerable financial subsidies and new regulation to promote hydrogen infrastructure investments in the next years, energy policy research is needed to guide such policies with holistic analyses. In this study, we link a electrolytic hydrogen supply chain model with an electricity system dispatch model, for a cross-sectoral case study of Germany in 2030. We find that hydrogen infrastructure investments and their effects on the electricity system are strongly influenced by electricity prices. Given current uniform prices, hydrogen production increases congestion costs in the electricity grid by 17\%. In contrast, passing spatially resolved electricity price signals leads to electrolyzers being placed at low-cost grid nodes and further away from consumption centers. This causes lower end-use costs for hydrogen. Moreover, congestion management costs decrease substantially, by up to 20\% compared to the benchmark case without hydrogen. These savings could be transferred into according subsidies for hydrogen production. Thus, our study demonstrates the benefits of differentiating economic signals for hydrogen production based on spatial criteria.
\end{abstract}


\maketitle

\section*{Nomenclature}
    \label{tab:notation_hydrogen_model}
    \begin{supertabular}{L{2.2cm}L{5.6cm}}
        &\\
        \multicolumn{2}{l}{\textbf{Decision variables}}\\
        $X_p\in\{0, 1\}$ & Hydrogen production/import at location p (1), or not (0)\\
        $HP_p\in[0,\infty)$ & Daily amount of hydrogen production at p [$kg_{H_2}$/day]\\
        $Y_{p,c}\in\{0, 1\}$ & Hydrogen transport from p to c (1), or not (0)\\
        $HT_{p,c}\in[0,\infty)$ & Daily amount of hydrogen transportation from p to c [$kg_{H_2}$/day]\\
        &\\
        \multicolumn{2}{l}{\textbf{Objective function parameters}}\\
        $PCC_p$ & Production capital cost at $p$ [EUR]\\
        $POC_p$ & Production operating cost at $p$ [EUR]\\
        $CCC_{p, s}$ & Conversion capital cost of $s$ at $p$ [EUR]\\
        $COC_{p, s}$ & Conversion operating cost of $s$ at $p$ [EUR]\\
        $TCC_s$ & Transportation capital cost of $s$ [EUR]\\
        $TOC_{s}$ & Transportation operating cost of $s$ [EUR]\\
        $SCC_s$ & Total fueling station capital cost of $s$ [EUR]\\
        $SOC_s$ & Total fueling station operating cost of $s$ [EUR]\\   
        \end{supertabular}
        \begin{supertabular}{L{4cm}L{3.8cm}}
        \multicolumn{2}{l}{\textbf{Indices}}\\
        $C$ & Set of consumption locations\\
        $P = P_{Production} \bigcup P_{Import}$ & Set of domestic production locations and import locations\\
		$S \in \{LH2, GH2, LOHC\}$ & Set of hydrogen transportation states\\
        &\\
        \multicolumn{2}{l}{\textbf{Exogenous parameters}}\\
        $a$ & Depreciation period [years]\\
        $AF$ & Annuity factor [\%]\\
        $CAP_{Import}$ & Import capacity [$kg_{H_2}$/day]\\ 
        $CAP_{Production, max}$ & Maximum production capacity [$kg_{H_2}$/day]\\ 
        $CAP_{Production, min}$ & Minimum production capacity [$kg_{H_2}$/day]\\ 
        $CAP_{Trailer_s}$ & Capacity of delivery trailer for state $s$ [$kg_{H_2}$]\\
        $DIST_{p,c}$ & Air-line distance between p and c [km] \\
        $DF$ & Detour factor [-]\\
        $DS$ & Driving speed [$km/h$]\\
        $DT_s$ & Duration of driving time [h]\\
        $EC$ & Electricity consumption [$kWh_{el}/kg_{H_2}$] \\
        $ED$ & Energy density of hydrogen [$kWh_{H_2}/kg_{H_2}$] \\
        $EE$ & Electric efficiency [$kWh_{H_2}/kWh_{el}$] \\
        $EP$ & Uniform single electricity price [EUR/$kWh_{el}$] \\
        $EP_{p}$ & Electricity price at p [EUR/$kWh_{el}$] \\
        $FC$ & Fuel consumption of delivery truck [$liter/km$] \\
        $FLH_{Electrolyzer}$ & Full load hours of electrolyzers [h] \\
        $FP$ & Fuel price [EUR$/liter$] \\
        $NGC_{Station_s}$ & Natural gas consumption of fuel station of hydrogen state s [$kWh_{NG}/kg_{H_2}$] \\
        $HD_c$ & Daily hydrogen demand at location c [$kg_{H_2}$/day]\\
        $HIC$ & Hydrogen import costs [EUR$/kg_{H_2}$]\\
        $IC_{Conversion_s}$ & Investment costs of conversion equipment [EUR] \\
        $IC_{Electrolyzer}$ & Capacity-dependent investment costs of electrolyzer [EUR/$kW_{el}$] \\
        $IC_{Station_s}$ & Investment cost per fuel station for state s [EUR] \\
        $IC_{Trailers_s}$ & Investment cost per trailer for state s [EUR] \\
        $IC_{Trucks}$ & Investment cost per truck [EUR] \\
        $LC$ & Labor costs\\
        $LT_s$ & Duration of loading and unloading for state s [h]\\
        $M$ & A very large positive number [-] \\
        $NS$ & Number of fuel stations [-]\\
        $NGP$ & Natural gas price [EUR/$kWh_{NG}$] \\
        $NT_s$ & Number of trucks for product state s [-] \\
        $O\&M$ & Operation and maintenance cost factor [\%]\\
        $T$ & Toll [EUR$/km$] \\
        $W$ & Wage [EUR$/hour$] \\
        $WACC$ & Weighted average cost of capital [\%] \\
    \end{supertabular}

\section{Introduction}
Hydrogen produced from low-carbon sources can contribute substantially to mitigating emissions in sectors that are difficult or impossible to electrify directly. Governments worldwide, and in particular in Europe, have announced strategies and billions of public funding to develop large-scale hydrogen infrastructure, that is centered on electrolytic hydrogen supply \citep{HydrogenCouncil.2021}. Since hydrogen production from electrolysis uses large amounts of electricity, a future hydrogen sector will introduce new interdependencies with the electricity sector. While electricity prices influence the cost-minimal installation \citep{vomscheidt.2021} and operation \citep{GUERRA20192425} of electrolyzers, these electrolyzers in turn introduce new electricity demand into the bulk power system, influencing in the short term the usage of renewable energy \citep{Ruhnau2020Market, BODAL202032899}, as well as congestion of power networks \citep{vomscheidt.2021, XIONG2021116201}, and in the long term the need for electricity generation and transmission capacity \citep{BODAL202032899}. Most importantly, the effects of these interdependencies will be strong and will prevail for a long time, because electrolyzers are large-scale, stationary consumers with typical lifetimes of ten years and more \citep{Schmidt2017}.\\

The integration of electrolyzers in European grids raises some unique questions as European wholesale power markets are designed as single-price zonal markets that overlook intra-zonal transmission capacities and price variations. Such single-price zonal market designs are already leading to rising congestion management costs in many electricity systems \citep{Staudt2017}. In Germany, the costs for congestion management have risen to almost a billion Euro annually, and especially the curtailment of renewable energy plants is increasing \citep{XIONG2021116201}. Yet, political decision makers have made it clear that nodal pricing will not be introduced in Germany, or Europe, any time soon \citep{Koalitionsvertrag.2018, EuropeanNetworkofTransmissionSystemOperatorsforElectricity.2021}. Without appropriate policy to guide system-beneficial integration, hydrogen production might strongly aggravate these effects. While the importance of market cost-reflective price-regulation and subsidization of electrolyzers has been voiced in the political sphere \citep{EU_hydrogen_strategy}, there is a prevailing lack of energy policy research to guide efficient integration of hydrogen infrastructure into the electricity sector.\\

Therefore, in this study, we link an electrolytic hydrogen supply chain model with an electricity system dispatch model to analyze the cost-minimal hydrogen infrastructure setup under different electricity price signals, using Germany as a case study. We find that under current regulation with uniform electricity prices, the cost-minimal solution is to produce hydrogen close to locations of consumption. These locations partly coincide with high locational marginal electricity costs. Consequently, our results also show how hydrogen production aggravates the inefficiencies of single-price markets and increases congestion management costs substantially.\\

We compare this benchmark scenario to a case in which electrolyzers are offered dedicated nodal tariffs, based on the locational marginal prices that would form in a nodal pricing system. We find that such nodal signals lead to higher shares of hydrogen production at low-price nodes, longer transport distances, and lower total costs for hydrogen. This demonstrates the sensitivity of hydrogen supply chains to spatial prices or subsidies. Moreover, in this scenario, the integration of hydrogen leads to congestion management costs that are substantially lower than in the benchmark scenario and even below the baseline scenario without hydrogen. Interestingly, these avoided redispatch costs could already cover most of the subsidies a regulator would have to pay to mimic nodal prices for hydrogen within the existing single-price market design.\\

Thus, in a time in which many policy makers and regulators in single-price markets are planning future hydrogen supply systems, electricity tariff designs for electrolyzers, and subsidies for hydrogen infrastructure, our study demonstrates and quantifies the considerable benefits of differentiating these economic signals with respect to spatial criteria.


\section{Background}
    Several past studies have addressed the spatial aspects of hydrogen supply chains that use grid electricity for hydrogen production. \citet{Robinius_2017_Part2, Reuss2019, emonts2019} present models that link a hydrogen supply chain with a national electricity grid. 
    The authors apply their model to the case of hydrogen fueled passenger cars in Germany in 2050 and identify favorable regions for hydrogen production in Germany. The study does not explicitly consider effects of (spatial) economic signals, but rather takes a technical supply chain perspective. \cite{Runge2019} optimize supply chains for synthetic fuels, including hydrogen stored in liquid organic hydrogen carrier (LOHC) material. Besides considering uniform single-prices, the authors also present a case in which they calculate state-level representative nodal prices for two exemplary states in Germany (NUTS-2 level) and allow transportation of hydrogen between the two states. This causes increased hydrogen production in the state with lower prices. The authors acknowledge the importance of future work analyzing feedback effects on the electricity system. We fill their identified gap with this study. 
    
    \cite{ZHANG2020115651} analyze the flexible operation of electrolyzers that produce hydrogen for light, medium- and heavy-duty fuel cell electric vehicles (FCEVs) in the Western United States of America. They find evidence that increasing electrolyzer flexibility lowers hydrogen and electricity generation cost and $CO_2$ emissions. With a similar focus on temporal aspects, it has been demonstrated that flexibility of electrolytic hydrogen production enables more renewable integration for case studies in Texas, USA \citep{BODAL202032899} and Germany \citep{Ruhnau2020Market}.
    
    \cite{Rose2020} focus on hydrogen supply for heavy-duty trucks from on-site electrolysis at highway fuel stations. They jointly optimize the infrastructure of fuel stations and the electricity system. They find that using 100\% hydrogen fueled heavy-duty trucks in Germany in 2050 would increase the total electricity demand by about 60 TWh and cause additional infrastructure costs of about 12 billion euro per year. They note that nodal prices contain important information about ''cost-effective energy consumption from a system perspective'' and that investors in hydrogen infrastructure should consider the system perspective. This idea is expanded and implemented by \cite{vomscheidt.2021}. The authors link a hydrogen supply chain optimization model and a nodal electricity system dispatch model and observe their interdependence in an initial case study of hydrogen-fueled trucks and passenger cars. They find that compared to current uniform single-price prices, nodal prices would lead to more hydrogen generation at low-price nodes. This in turn causes substantially lower congestion management costs. However, their analysis, like all previous ones, focuses on a small subset of hydrogen demand, i.e. demand from the transport sector.
    
    \cite{XIONG2021116201} provide another perspective on the topic of hydrogen integration in single-price markets. They do not consider the effects of hydrogen production on day-ahead energy wholesale markets, but analyze how Power-to-Gas plants (i.e. electrolyzers) can serve as a redispatch option. They find that curtailment of renewable generation can be reduced by 12\% when electrolyzers are installed for performing redispatch at a few frequently curtailed nodes in the German grid of 2015. The study thus showcases the importance of spatial consideration in hydrogen infrastructure planning. However, the study disregards spatial aspects of the hydrogen supply chain and disregards how policy makers could incentivize investors to build electrolyzers at the identified nodes. Moreover, the direct political applicability of the study is restricted, because future hydrogen volumes, shares of renewable and conventional generation, and spatial distribution of generation will be much different than in the used scenario from 2015.
    
    In summary, past research indicates that the spatial dimension of hydrogen integration matters. Within the limitations of single sector analyses and/or a reduced network consideration, studies have demonstrated that electrolyzer locations influence grid congestion. However, to the best of our knowledge, no past study has evaluated the cost-optimal hydrogen supply chain for electrolytic hydrogen use for a broad range of hydrogen demand sectors, namely steel, ammonia, methanol, refineries, and transportation, considered the effect of alternative electricity price signals, and assessed the feedback effects of the resulting supply chains. Such analysis is timely from a policy perspective given the prospect of significant electrolyzer capacity integration over the next decade in the German power grid. Therefore, we pose the following research questions to guide future policy decision on hydrogen subsidization:
    \begin{enumerate}
        \item What is the cost-minimal supply chain design using electrolytic hydrogen production for the combined hydrogen demand from all major relevant sectors in 2030 in Germany?
        \item How do electricity price signals (single-price versus nodal) influence the cost-minimal locations of electrolyzers and the costs of hydrogen?
        \item How does hydrogen production change electricity wholesale prices and congestion management costs under single-price and nodal price signals?
    \end{enumerate}

\section{Methods} \label{Section:Methodology}
\begin{figure*}
    	\centering
    		\includegraphics[scale=.50]{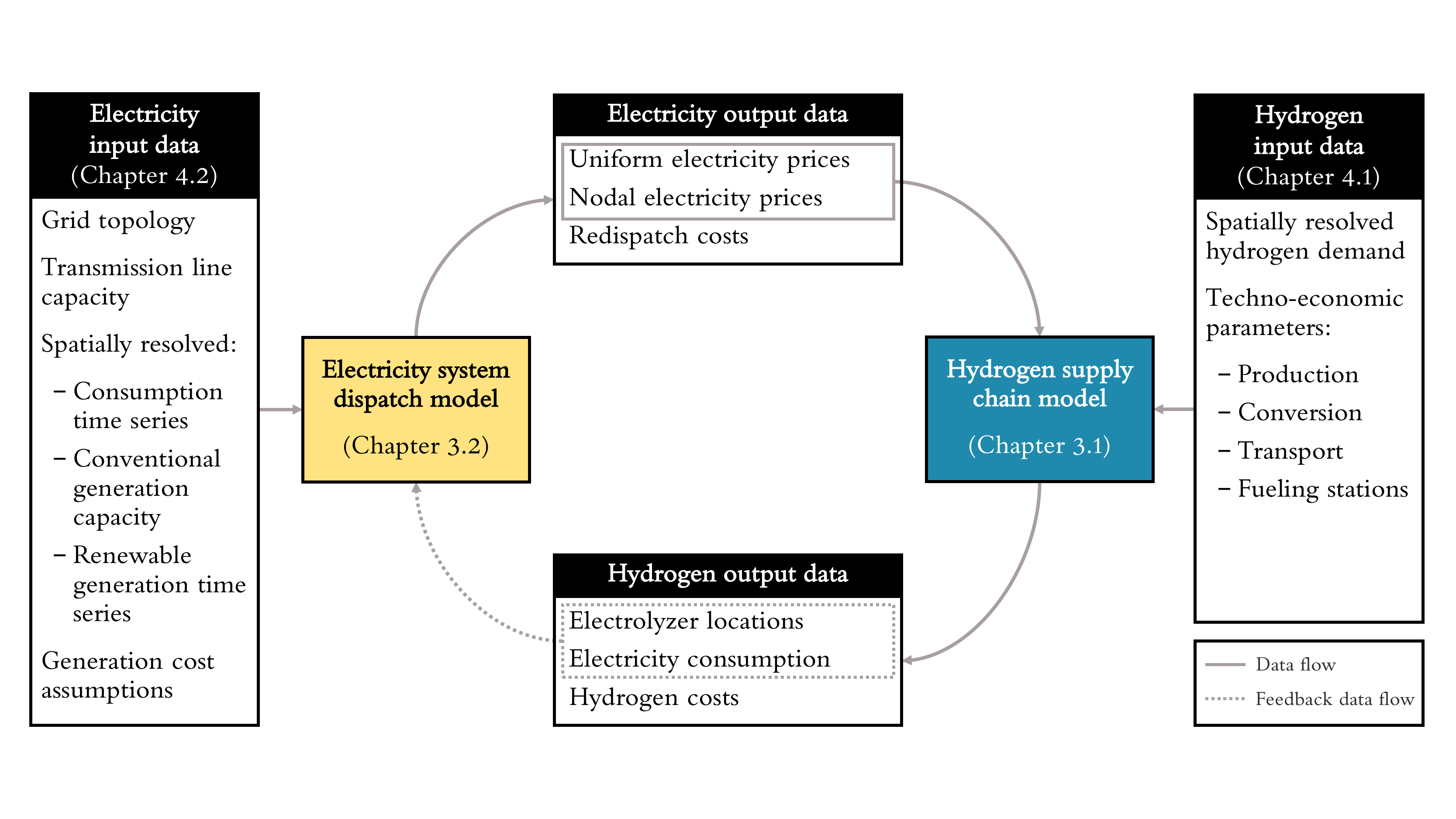}
    	\caption{Method overview: Models for the hydrogen and the electricity system are linked through inputs and outputs}
    	\label{fig:model}
    \end{figure*}

    To address the research questions, we model the hydrogen supply chain and the electricity system and link both models through their respective inputs and outputs. As shown in Figure \ref{fig:model}, we proceed in three steps. First, we parametrize an electricity system dispatch model without hydrogen and compute uniform prices, nodal prices, and redispatch costs. Second, utilizing the computed electricity prices, we run the hydrogen model to identify the cost-minimal spatial siting of electrolyzers, their capacities, and the hydrogen transportation.\footnote{Note that we neglect grid charges, taxes and other charges commonly included in an end-user tariff and assume the calculated electricity prices equal the final tariff payed by the electrolyzer operator.} We consider a scenario for uniform prices to reflect current regulation and the a scenarios for nodal prices, to reflect a more efficient solution. For both scenarios we calculate both a case of static electrolyzer operation, and a case of dynamic/flexible operation. Third, we feed back the resulting locations and capacities of the electrolyzers as additional regional loads into the electricity model. We calculate consequential changes in wholesale electricity prices and congestion management costs. Both models are implemented in Python 3.7.3, and solved using the Gurobi solver 8.1.1. 
    
    \subsection{Hydrogen supply chain model}
        In the following, we describe the details of the hydrogen supply chain model. It represents an enhanced version of the model in \citep{vomscheidt.2021}. Due to the temporal uncertainty in demand from end-use sectors the model assumes time-invariant hydrogen consumption.
        
    
        \subsubsection{Objective function}
            The model minimizes the total annual end-use costs of hydrogen, which consist of capital costs and operating costs for electrolytic production ($PCC$, $POC$), conversion ($CCC$,\linebreak $COC$) and transportation ($TCC$, $TOC$) of hydrogen, and, in the case of hydrogen use in the transportation sector, the fueling stations ($SCC$, $SOC$) (\ref{equ:total annual cost}). For this, the model optimizes the location and size of electrolyzers and the amount of hydrogen that is transported from each electrolyzer to each location of consumption.
            There are four decision variables. (i) $X_p$ is a binary variable that indicates whether an electrolyzer is installed at a location $p$ (1) or not (0). (ii) $HP_p\in[0,\infty)$ denotes the amount of hydrogen produced at $p$ in $kg_{H_2}$ per day. (iii) $Y_{p,c}$ is binary and indicates whether hydrogen is transported from a production location $p$ to a consumption location $c$ (1) or not (0). (iv) $HT_{p,c}\in[0,\infty)$ denotes the amount of hydrogen transported from $p$ to $c$ in $kg_{H_2}$ per day. $P$ and $C$ represent the set of all potential electrolyzer plant locations $p$, and consumption locations $c$, respectively. Thus, the model outputs the cost-minimal locations of electrolyzers, their individual daily production, the transportation volume between each electrolyzer and point of consumption, and the resulting end-use costs of hydrogen.
            \begin{equation}
            \label{equ:total annual cost}
            \begin{split} 
            \min_{X_p, HP_p, Y_{p,c}} (\sum_{p \in P} PCC_p(\boldsymbol{X_p, HP_p}) + \sum_{p \in P} POC_{p}(\boldsymbol{X_p, HP_p}) \\ + \sum_{p \in P} CCC_{p,s}  + \sum_{p \in P} COC_{p, s}(\boldsymbol{X_p, HP_p}) \\ + TCC_{s} + \sum_{p \in P}\sum_{c \in C} TOC_{p,c,s}(\boldsymbol{Y_{p,c}, HT_{p,c}}) \\ + SCC_{s} + SOC_{s})
            \end{split} 
            \end{equation}
            
            The model can be parametrized for three possible hydrogen states $s$ for transportation via delivery trailers: liquefied (LH2), compressed gaseous (GH2), and bound in LOHC. These three states require different technologies for conversion, transportation and fueling stations and thus cause different costs. The annotation of decision variables, indices and input variables is provided in the nomenclature.\\
            
            The four components of capital costs include specific annual operation and management costs ($O\&M$) and annuity factors ($AF$). The annuity factors account for the depreciation of one-time investments over multiple years and depend on the weighted average cost of capital ($WACC$ [\%]) and depreciation periods ($a$ [years]) (Eq. \ref{equ:AF}). 
            
            \begin{equation} 
            \label{equ:AF}
            AF = \frac{(1+WACC)^{a} * WACC}{(1+WACC)^{a}-1}
            \end{equation}
            
            Production capital costs occur at a location only if an electrolyzer is placed there, and depends on the daily hydrogen output capacity (Eq. \ref{equ:PCC}).
            \begin{equation} 
            \label{equ:PCC}
            \begin{split} 
            PCC_p = \boldsymbol{X_p} * \frac{\boldsymbol{HP_p} * ED * IC_{Electrolyzer}}{FLH * EE} \\ * (1 + O\&M_{Electrolyzer}) \\ * AF_{Electrolyzer} \hspace{2mm} \forall p \in P_{Production}
            \end{split}
            \end{equation}
            
            Production operating costs depend on the amount of hydrogen that is produced, the efficiency of the electrolyzer, and the electricity price $EP_p$ (Eq. \ref{equ:POC}). Note that this price varies with location p.
            \begin{equation}
            \label{equ:POC}
            \begin{split}
            POC_p = \boldsymbol{HP_{p}} * EC_{Production} * EP_{p} * 365 \hspace{2mm} \\ \forall p \in P_{Production}
            \end{split}
            \end{equation}
            
            In addition to domestic hydrogen production, the model also includes overseas imports. Imports do not incur any capital costs for production (Eq. \ref{equ:PCC_import}), but specific production operating costs (Eq. \ref{equ:POC_imports}).
            \begin{equation} 
            \label{equ:PCC_import}
            PCC_{Import} = 0
            \end{equation}
            \begin{equation}
            \label{equ:POC_imports}
            POC_{Import} = HP_{Import} * HIC * 365 
            \end{equation}
            
            Moreover, conversion capital costs are assumed to occur in bulk (depending on the total amount of hydrogen that needs to be converted daily) and independently of the location (Eq. \ref{equ:CCC}). They do however, depend on the state in which hydrogen is to be transported afterwards, namely gaseous, liquefied, or stored in LOHC.
            \begin{equation}
            \label{equ:CCC}
            \begin{split} 
            CCC_{s} = IC_{Conversion_s} * (1 + O\&M_{Conversion_s}) \\ * AF_{Conversion_s} \hspace{2mm} \forall s \in S
            \end{split} 
            \end{equation}
            
            Furthermore, the model includes the operating costs of converting hydrogen. 
            For liquid delivery, hydrogen needs to be liquefied and later evaporated at the location of consumption, which requires electricity (Eq. \ref{equ:COC_LH2}). 
            \begin{equation} 
            \label{equ:COC_LH2}
            \begin{split} 
            COC_{p, LH2} = (\sum_{p \in P_{Production}} \boldsymbol{HP_p} * EC_{Liquefaction} \\ * EP_p  * (1 + Loss_{Liquefaction})) \\ + (\sum_{p \in P} \boldsymbol{HP_p} * EC_{Evaporation} * EP \\ * (1 + Loss_{Evaporation})) * 365
            \end{split}
            \end{equation}
            
            In the case of gaseous delivery, hydrogen is compressed. This causes  (Eq. \ref{equ:COC_GH2}). 
            \begin{equation} \label{equ:COC_GH2}
            \begin{split}
            COC_{p, GH2} = \sum_{p \in P_{Production}} \boldsymbol{HP_p} * EC_{Compression} \\ * EP_p * (1 + Loss_{Compression}) * 365
            \end{split}
            \end{equation}
            
            For LOHC delivery, the carrier material needs to be hydrogenated and later dehydrogenated at the location of consumption (Eq. \ref{equ:COC_LOHC}). Electricity is required for both steps. Additionaly, natural gas is required for dehydrogenation.
            \begin{equation} 
            \label{equ:COC_LOHC}
            \begin{split} 
            COC_{p, LOHC} = (\sum_{p \in P \setminus Import} \boldsymbol{HP_p} * EC_{Hydrogenation} \\ * EP_p * (1 + Loss_{Hydrogenation})) \\ + (\sum_{p \in P} \boldsymbol{HP_p} * (EP * EC_{Dehydrogenation} \\ + NGP * NGC_{Dehydrogenation}) \\ * (1 + Loss_{Dehydrogenation})) * 365
            \end{split}
            \end{equation}
            
            For LH2 and LOHC, we assume that imports already arrive in the respective form and thus do not require the first conversion step for domestic delivery.
            After initial conversion, hydrogen is transported to the consumption sinks. In general, hydrogen can be transported via tube trailers mounted onto delivery trucks, or via pipelines. Since related work indicates that transport via pipelines only becomes economically viable for long transport distances in high demand scenarios \citep{Reuss2019, Robinius_2017_Part2, TLILI.2020}, our model focuses on transport via tube trailers on delivery trucks.\footnote{Future work could expand our model by including both truck based and pipeline based hydrogen transportation. This could lead to lower end-use costs. However, it would likely not affect this study’s findings regarding optimal electrolyzer locations and redispatch in a substantial manner, because hydrogen transportation costs have a much smaller impact on total costs and thus optimal locations than hydrogen production costs (compare \ref{fig:cost_components}). Therefore, even if transportation costs were lower in a pipeline scenario, this would not lead to different results regarding cost-optimal location of electrolyzers and redispatch.} Thus, transport capital costs depend on the investment costs for hydrogen trailers and the respective transport trucks, as well as the number of trailers and trucks. Each truck carries one trailer.
            \begin{equation} 
            \label{equ:TCC}
            \begin{split} 
            TCC_{s} = IC_{Trucks} * NT_s * (1 + O\&M_{Trucks}) \\ * AF_{Trucks} + IC_{Trailers_s} * NT_s \\ * (1 + O\&M_{Trailers_s}) * AF_{Trailers} \hspace{2mm}\forall s \in S
            \end{split}
            \end{equation}
            
            Transport operating costs consist of costs for labor, fuel, and toll (Eq. \ref{equ:TOC}). 
            \begin{equation} \label{equ:TOC}
            TOC_s = LC + FC + TC
            \end{equation}
            
            Labor costs depend on the time that drivers spend loading, driving, and unloading the delivery trailers (Eq. \ref{equ:LC}). A fixed loading and unloading time per delivery is assumed. 
            \begin{equation} \label{equ:LC}
            LC = (DT_s + LT_s * NT_s) * W
            \end{equation}
            
            Round-trip driving time is determined by the distance between connected production plants and points of consumption, as well as the driving speed $DS$. Transport distances are approximated via air-line distance, multiplied with a detour factor of 1.3, in line with \cite{Reuss2019Dissertation}. Since the daily capacity of fueling stations is assumed to be smaller than the capacity of one delivery trailer, we multiply the distances to fueling stations with a frequency factor $HD_c / CAP_{Trailer_s}$ (Eq. \ref{equ:drivingtime}) simulating that they are not provided with hydrogen on a daily basis. This also applies to fuel and toll costs (Eq. \ref{equ:FTC}).
            \begin{equation} \label{equ:drivingtime}
            \begin{split}
            DT_s = \frac{2 * DF}{DS} * (\sum_{p \in P} \sum_{c \in C_{Industry}}{\boldsymbol{Y_{p,c}} * DIST{p,c}} \\ + \sum_{p \in P} \sum_{c \in C_{Stations}}{\boldsymbol{Y_{p,c}} * DIST_{p,c}} * \frac{HD_{c}}{CAP_{Trailer_s}})
            \end{split} 
            \end{equation}
            \begin{equation} \label{equ:FTC}
            \begin{split}
            FTC = 2 * (FC_{Truck} * FP + T) * DF \\ * ((\sum_{p \in P} \sum_{c \in C_{Industry}}{\boldsymbol{Y_{p,c}} * DIST{p,c}}) \\ + (\sum_{p \in P} \sum_{c \in C_{Stations}}{\boldsymbol{Y_{p,c}} * DIST_{p,c}} * \frac{HD_{c}}{CAP_{Trailer_s}}))
            \end{split}
            \end{equation}

            For hydrogen that is to be used to refuel fuel cell trucks or passenger cars one additional supply chain step is modelled: the fuel station. The capital costs for fuel stations depend on the investment costs per station and the total number of stations (Eq. \ref{equ:SCC}).
            \begin{equation} 
            \label{equ:SCC}
            \begin{split} 
            SCC_{s} = {IC_{Station_s} * NS} * (1 + O\&M_{Station}) \\* AF_{Station} \hspace{2mm} \forall s \in S
            \end{split} 
            \end{equation}
            
            The operating costs depend on the required consumption of electricity and natural gas, and their respective prices (Eq. \ref{equ:SOC}).
            \begin{equation} \label{equ:SOC}
            \begin{split}
            SOC_{s} = (EC_{Station_s} * EP + NGC_{Station_s} * NGP) \\ * (1 + Loss_{Station_s}) * \sum_{c \in C_{Stations}} HD_{c} * 365
            \end{split}
            \end{equation}
        
        \subsubsection{Constraints}
            The model includes both domestic production and an exogenously given import at one fixed node. The sum of daily domestic and imported hydrogen production $HP$ must satisfy the sum of the exogenously given demand $HD$ (Eq. \ref{equ:HD}). 
            \begin{equation} \label{equ:HD}
            \sum_{p \in P} \boldsymbol{HP_{p}} = \sum_{c \in C} HD_{c}
            \end{equation}
            
            Hydrogen output $HP_p$ of each electrolyzer depends on its installed capacity, which varies between a fixed minimum (Eq. \ref{equ:ProductionCapaMin}) and maximum (Eq. \ref{equ:ProductionCapaMax}) value.
            \begin{equation} \label{equ:ProductionCapaMin}
            \boldsymbol{HP_{p}} \geq CAP_{Production, min} * \boldsymbol{X_{p}} \hspace{2mm} \forall p \in P_{Production}
            \end{equation}
            
            \begin{equation} \label{equ:ProductionCapaMax}
            \boldsymbol{HP_{p}} \leq CAP_{Production, max} * \boldsymbol{X_{p}} \hspace{2mm} \forall p \in P_{Production}
            \end{equation}
            
            The import nodes and their capacity are exogenously set in Eq. \ref{equ:ProductionBinary_Import} and \ref{equ:ProductionCapa_Import}. 
             \begin{equation} \label{equ:ProductionBinary_Import}
            \boldsymbol{X_{p}} = 1 \hspace{2mm} \forall p \in P_{Import} 
            \end{equation}
            \begin{equation} \label{equ:ProductionCapa_Import}
            \boldsymbol{HP_{p}} = CAP_{Import} * \boldsymbol{X_{p}} \hspace{2mm} \forall p \in P_{Import}
            \end{equation}
            
            In sum, the daily amount of hydrogen transported from a hydrogen source (electrolyzer or import) must not exceed its production (Eq. \ref{equ:Transport<Production}). 
            \begin{equation} 
            \label{equ:Transport<Production}
            \sum_{c \in C}{\boldsymbol{HT_{p, c}}} \leq \boldsymbol{HP_p} \hspace{2mm} \forall p \in P
            \end{equation}
            
            The daily amount transported to a consumer must meet its demand (Eq. \ref{equ:Transport>Demand}). 
            \begin{equation} 
            \label{equ:Transport>Demand}
            \begin{split} 
            \sum_{p \in P}{\boldsymbol{HT_{p, c}}} \geq HD_c \hspace{2mm} \forall c \in C
            \end{split}
            \end{equation}
            
            Positive transport volume from a plant to a consumption location is only possible if the delivery connection is established via binary variable $Y$ (Eq. \ref{equ:Transport_binary_volume}). 
            \begin{equation} 
            \label{equ:Transport_binary_volume}
            \begin{split} 
            \boldsymbol{HT_{p,c}} = 0, if \hspace{2mm} \boldsymbol{Y_{p,c}} = 0 \hspace{2mm} \forall c \in C, \forall p \in P\\
            \boldsymbol{HT_{p,c}} > 0, if \hspace{2mm} \boldsymbol{Y_{p,c}} = 1 \hspace{2mm} \forall c \in C, \forall p \in P
            \end{split} 
            \end{equation}
            
            Thus, one limitation of the model is that it does not consider short term or long term temporal variations in hydrogen transportation or consumption and thus neglects storage. While out of scope of this study, future work could attempt to identify short term and long term temporal patterns of hydrogen demand from industry and transportation.\footnote{Regarding long term storage, techno-economic parameters are presented by \citet{Reuss2019} and locations with high geological potential for hydrogen storage are presented by \cite{CAGLAYAN20206793}.} We do assess a sensitivity case "FlexOp" in which transportation and consumption remain continuous, but the production is temporally flexible. This allows us to identify an optimistic estimate of the potential cost savings that flexible electrolyzer operation can yield.
            
        \subsection{Electricity system model}
        Next, we model the electricity system to calculate electricity prices and congestion management costs, following the approach introduced in \citet{vomscheidt.2021}.
    
        For the uniform price scenario, we adapt a stylized merit-order and redispatch model from \cite{Staudt2020}. For each hour, the model minimizes the marginal generation costs for the entire single-price market area (Eq. \ref{equ:uniform_model_1}). The model's constraints ensure that demand and supply are balanced (Eq. \ref{equ:uniform_model_2}) subject to the constraints that limit available generation capacity (Eq. \ref{equ:uniform_model_3}). The annotation is given in Table \ref{tab:notation_electricity_model_parameters}.
        \begin{equation}
        \label{equ:uniform_model_1}
        \min_{}{(\sum_{t = 1}^T\sum_{n = 1}^N\sum_{g = 1}^G{q_{n,g,t} * p_{n,g}})}
        \end{equation}
        \begin{equation}
        \label{equ:uniform_model_2}
        \sum_{n = 1}^N{d_{g,t}} = \sum_{n = 1}^N\sum_{g = 1}^G{q_{n,g,t}} \hspace{2mm} \forall t \in T
        \end{equation}
        \begin{equation}
        \label{equ:uniform_model_3}
        \hspace{0.5cm}q_{n,g,t} \leq c_{n,g,t} \hspace{2mm} \forall g \in G \hspace{2mm} \forall n \in N, \hspace{2mm} \forall t \in T
        \end{equation}
                 
        Complying with the market designs of single-price markets, this model does not consider grid constraints. Therefore, the resulting market allocation can be technically infeasible, in which case redispatch measures ensue, modelled by Eq. \ref{equ:redispatch_model_1} to \ref{equ:redispatch_model_5}. The cost-based redispatch mechanism begins at the existing market allocation and activates and deactivates generation capacity in the system until the cost-minimal solution is found that respects grid constraints which in the optimal case is equivalent to the nodal pricing solution \citep{staudt2019transmission}. Generators that are activated through this procedure are compensated based on their operating costs. The additional costs caused by this procedure are referred to as redispatch costs. In the considered idealized case, they are equivalent to the congestion management costs.
        \begin{equation}
        \label{equ:redispatch_model_1}
        \min_{}{(\sum_{n = 1}^N \sum_{g = 1}^G{q^{\Delta}_{n,g,t} * p_{n,g}})} \hspace{2mm} \forall t \in T
        \end{equation}
        \begin{equation}
        \label{equ:redispatch_model_2}
        \sum_{n = 1}^N \sum_{g = 1}^G{q^{\Delta}_{n,g,t} = 0} \hspace{2mm} \forall t \in T
        \end{equation}
        \begin{equation}
        \label{equ:redispatch_model_3}
        q^{\Delta}_{n,g,t} + q_{n,g,t} \leq c_{n,g, t} \hspace{2mm} \forall n \in N, \hspace{2mm} \forall g \in G, \hspace{2mm} \forall t \in T
        \end{equation}
        \begin{equation}
        \label{equ:redispatch_model_4}
        q^{\Delta}_{g,t} + q_{g,t} \geq 0 \hspace{2mm} \forall n \in N, \hspace{2mm} \forall g \in G, \hspace{2mm} \forall t \in T
        \end{equation}        
        \begin{equation}
        \label{equ:redispatch_model_5}
        \begin{split}
        |\sum_{n = 1}^N{\sum_{g = 1}^G{(((q_{n,g,t} + q^{\Delta}_{n,g,t}})} - d_{n, t}) * H_{l,n}))| \hspace{2mm} \leq \tau_{l} \\ \forall l \in L, \hspace{2mm} \forall t \in T
        \end{split}
        \end{equation}

        For the nodal price scenario, we use a nodal model with a DC-load flow approximation. This model simultaneously takes into account generation capacities and prices (Eq. \ref{equ:uniform_model_1} to \ref{equ:uniform_model_3}), as well as transmission capacities (Eq. \ref{equ:nodal_model}).
        \begin{equation}
        \label{equ:nodal_model}
        \begin{split}        
        |\sum_{n = 1}^N{\sum_{g = 1}^G{((q_{g,t}}} - d_{n, t}) * H_{l,n})| \leq \tau_{l} \hspace{2mm} \forall l \in L, \hspace{2mm} \forall t \in T
        \end{split}
        \end{equation}
        
        Both models optimize each hour step-wise, independently of other hours. They thus neglect generation ramping and storage.
        
        \begin{table}[width=1.0\linewidth,cols=2,pos=htbp]
           \caption{Notation for electricity system model} 
            \label{tab:notation_electricity_model_parameters}
            \begin{tabular}{L{0.6cm}L{6.7cm}}
                \toprule
                    $q_{n,g,t}$& Generation of unit g at node n at time t\\
                    $q^{\Delta}_{n,g,t}$ & Redispatch of unit g at node n at time t\\
                    $p_{n,g}$& Marginal generation costs of unit g at node n\\
                    $d_{n,t}$& Demand at node n at time t\\
                    $c_{n,g,t}$& Generation capacity of unit g at node n (at time t for renewables)\\
                    $\tau_{l}$& Transmission capacity of line l \\
                    $H$& Matrix of power distribution factors\\
                    $N$& Number of nodes n\\
                    $G$& Number of generation units g\\
                    $L$& Number of lines l\\
                \bottomrule
            \end{tabular}
            \end{table} 
        
\section{Case study} \label{Section:Case_Study}
    To demonstrate the functioning of the hydrogen model and the electricity model, we apply it to a case study. For this, we parametrize the models with data for the German electricity system and hydrogen demand in 2030.

    \subsection{Hydrogen data}
    In this subsection, we present all data sources, preprocessing steps, and assumptions used for creating the input data sets for demand, production, conversion, and transportation of hydrogen.
    
        \subsubsection{Hydrogen demand}
            In the following paragraphs, we describe data acquisition and preprocessing for the German hydrogen net demand in 2030. Hydrogen demand is assumed to come from the six following sectors: steel, ammonia, methanol, refinery, road transportation, and individual mobility. First, each demand sector is presented with general assumptions about future hydrogen demand and potential. Subsequently, the relevant locations of the respective sector with hydrogen demand in 2030 are identified. The estimated total hydrogen net demand amounts to 51.26 TWh. Table \ref{tab:hydrogen_conversion_factors} in appendix \ref{appendix_B} shows the numeric values and conversion factors used for the hydrogen demand calculations. For steel, ammonia, methanol, and refineries, 100\% availability of the production facilities is assumed. Correspondingly, quantities that have been calculated down to hours are multiplied by 8,760 to get the respective annual quantity. For details on data acquisition and processing we refer to appendix \ref{appendix_A}.
            
            \paragraph{\textbf{Ammonia}}
            Ammonia (NH\textsubscript{3}) is produced using the Haber-Bosch process and requires the input components hydrogen (H\textsubscript{2}) and nitrogen (N\textsubscript{2}) \citep{oei_874}.
            The potential for CO\textsubscript{2} emissions reduction lies in replacing fossil-generated hydrogen with electricity-based hydrogen. Today, hydrogen is mostly produced from steam methane reforming, with the by-product CO\textsubscript{2}. This byproduct can be used for processes in material composites, such as the production of urea \citep{hebling2019wasserstoff}. Nevertheless, our estimation assumes a complete switch of ammonia production to electricity-based hydrogen in order to define an upper limit of hydrogen demand in the ammonia industry. Table \ref{tab:ammonia} summarises the hydrogen demand of the ammonia industry. Based on the assumptions made, the total hydrogen demand is 17.49 TWh, distributed over four plants.

            \begin{table}[htbp]
            \caption{Estimated hydrogen demand of ammonia producers in Germany, 2030}
            \label{tab:ammonia}
            \begin{tabular}{@{}ll@{}}
            \toprule
            Ammonia producer & \begin{tabular}[c]{@{}l@{}}Hydrogen net \\ demand {[}TWh{]}\end{tabular} \\
            \midrule
            BASF Ludwigshafen & 5.18 \\
            INEOS Köln & 2.25\\
            SKW Stickstoffwerke Piesteritz & 5.62 \\
            YARA Brunsbüttel & 4.44 \\ 
            \midrule
            Total & 17.49 \\ 
            \bottomrule                                                                  
            \end{tabular}
            \end{table}
            \paragraph{\textbf{Steel}}
            Steel production in Germany offers a large potential for the use of hydrogen in industry by switching to hydrogen-based processes. In general, a distinction is made in steel production between primary and secondary steel as well as between blast furnace and electric arc routes \citep{hebling2019wasserstoff}. Today, primary steel production is mainly based on coal- or coke-based processes to reduce iron ore in the blast furnace, resulting in large amounts of carbon emissions \linebreak \citep{wilms2018heutige}. An alternative to the blast furnace is direct reduction, in which the iron ore is reduced by natural gas or hydrogen and CO\textsubscript{2} emissions are directly avoided \citep{hebling2019wasserstoff}. The product "direct reduced iron" is further processed into steel in an electric arc furnace. If hydrogen is produced by electrolysis with electricity from renewable energies and used instead of coal in the direct reduction process, up to 95 \% of CO\textsubscript{2} emissions could be avoided on the way to primary steel \citep{Berger.2020}. In addition to the possibility of switching to direct reduction, CO\textsubscript{2} emission reductions can be achieved by blowing in hydrogen as a substitute reducing agent. The basic idea is to reduce the amount of injection coal required and to replace it with hydrogen, in order to reduce CO\textsubscript{2} emissions \citep{thyssenkrupp.2019}. Depending on the operating conditions, emissions can be reduced by 21.4 - 28.5 \% compared to a reference case with today's standard operating mode \citep{yilmaz2018massnahmen}.
            
            We identify all steel plants with potential for hydrogen use in 2030 through an extensive review of industry reports and press releases, as elaborated in \ref{appendix_A}. Table \ref{tab:steel_demand} summarizes the hydrogen net demand of the steel industry. Based on the assumptions made, the total hydrogen net demand for 2030 amounts to 13.25 TWh and is distributed over Hamburg, Dillingen/Saar, Peine and Duisburg.\\
            \begin{table}[htbp]
            \caption{Estimated hydrogen net demand of primary steel producers in Germany, 2030}
            \label{tab:steel_demand}
            \begin{tabular}{@{}ll@{}}
            \toprule
            Steel producer    & \begin{tabular}[c]{@{}l@{}}Hydrogen net \\ demand {[}TWh{]}\end{tabular}\\ 
            \midrule
            ArcelorMittal Bremen             & 0.0                \\
            ArcelorMittal Duisburg      & 0.0                \\
            ArcelorMittal Eisenhüttenstadt  & 0.0                \\
            ArcelorMittal Hamburg           & 2.67             \\
            ROGESA (Dillinger \& Saarstahl)  & 2.16             \\
            HKM Duisburg       & 0.0                \\
            Salzgitter Peine     & 2.25             \\
            Thyssenkrupp Steel Europe Duisburg & 6.17             \\ 
            \midrule
            Total  & 13.25 \\ 
            \bottomrule
            \end{tabular}
            \end{table}

            \paragraph{\textbf{Methanol}}
            Currently, methanol is commonly produced using synthesis processes with CO\textsubscript{2} emissions, which in future can be switched to hydrogen based processes \citep{michalski2019wasserstoffstudie}. Table \ref{tab:methanol} summarises the hydrogen demand of the methanol industry. Based on the assumptions made, the total hydrogen demand is 11.73 TWh and is distributed over four sites.
            
            \begin{table}[htbp]
            \caption{Estimated hydrogen net demand of methanol producers in Germany, 2030}
            \label{tab:methanol}
            \begin{tabular}{@{}lr@{}}
            \toprule
            Methanol producer & \multicolumn{1}{l}{\begin{tabular}[c]{@{}l@{}}Hydrogen net \\ demand {[}TWh{]}\end{tabular}} \\ 
            \midrule
            BASF Ludwigshafen      & 2.83            \\
            Shell Rheinland Raffinerie - Süd         & 2.74            \\
            Ruhr Oel - BP Gelsenkirchen & 1.76            \\
            Total Raffinerie Mitteldeutschland  & 4.40            \\ 
            \midrule
            Total & 11.73  \\ 
            \bottomrule
            \end{tabular}
            \end{table}
            
            \paragraph{\textbf{Refineries}}
            In refineries, hydrogen is used on a large scale to desulfurize fuels and to refine heavy residues with hydrogen via hydrocracking \citep{oei_874}. The hydrogen needed for crude oil processing is supplied from internal and external sources. This means that refineries are partly self-sufficient, since hydrogen is a by-product of other processing operations \citep{ENCON.EuropeGmbH.2018}. In this study, a 22 \% net demand for hydrogen is assumed, analogous to \cite{wilms2018heutige}. This hydrogen net demand is assumed to be entirely served by electricity-based hydrogen in 2030, in line with \cite{PrognosAG.2020}. Table \ref{tab:refinery} summarises the hydrogen net demand of the refineries in Germany 2030. The estimated total hydrogen net demand is 4.29 TWh and is distributed over thirteen sites in Germany.
            
            \begin{table}[htbp]
            \caption{Estimated hydrogen net demand of refineries in Germany, 2030}
            \label{tab:refinery}
            \begin{tabular}{@{}lr@{}}
            \toprule
            Refinery & \multicolumn{1}{l}{\begin{tabular}[c]{@{}l@{}}Hydrogen net \\demand {[}TWh{]}\end{tabular}} \\ 
            \midrule
            Bayernoil Raffineriegesellschaft & 0.19  \\
            BP Raffinerie Lingen & 0.21   \\
            Gunvor Raffinerie Ingolstadt  & 0.22  \\
            Holborn Europa Raffinerie  & 0.23   \\
            MiRO Mineraloelraffinerie Oberrhein & 0.66   \\
            Nynas & 0.08   \\
            OMV Deutschland  & 0.16   \\
            PCK Raffinerie  & 0.51   \\
            Raffinerie Heide  & 0.19   \\
            Ruhr Oel  - BP Gelsenkirchen &  0.57   \\
            Shell Rheinland Raffinerie Werk Nord & 0.41   \\
            Shell Rheinland Raffinerie Werk Süd &  0.32   \\
            Total Raffinerie Mitteldeutschland  & 0.53   \\ 
            \midrule
            Total & 4.29 \\ 
            \bottomrule
            \end{tabular}
            \end{table}
            
            \paragraph{\textbf{Transportation sector}}
            
            In the first step, we estimate total national hydrogen demand in the transport sector and the number of fueling stations required to satisfy the demand. For this, we calculate a main scenario with fuel cell trucks, and a sensitivity scenario with additional fuel cell passenger cars. In the second step, we spatially disaggregate this total demand and determine potential sites for fueling stations.
                    
            To determine the hydrogen demand for fuel cell trucks and passenger cars in Germany in 2030, we use the mean estimates from \cite{FraunhoferISE2019}, namely 1 TWh for trucks and 3 TWh for cars. We assume that \textit{heavy-duty} trucks with a total weight above 12,000 kg \citep{VehicleDefinition} will be responsible for all truck based demand, because they have particularly high carbon emission savings potential and the fuel cell based version has stronger advantages over to their battery based counterparts, i.e. heavier payloads, longer ranges, and shorter recharging times \citep{WEGER2020}.
            We assume the consumption of trucks to decrease to 8 kg/100km until 2030, and that of fuel cell passenger cars to decrease to 0.63 kg/100km, in line with \cite{Grube2018, FCH-JU2017}; \cite{Hyundai2020}.\footnote{This translates to ca. 1 million fuel cell cars (2.6 \% penetration rate), and ca. 11,000 fuel cell trucks (2,4 \% penetration rate.}
            We assume that by 2030 all hydrogen stations will become L-size \citep{H2StationStandard} with a capacity of 1,000 kg/day. According to \cite{Reuss2019}, station investment cost is estimated considering scaling and learning effects, based on Equation (\ref{equ:Station investment}). With the total number of fuel stations (n) determined in our model, a capacity of each fuel station C = 1,000 kg/day, and the exogenous parameters $\alpha$, $\beta$, and $\gamma$ presented in Table \ref{tab:Hydrogen station assumptions}, we derive the investment cost per station for each hydrogen transportation state $s$.
            \begin{equation}\label{equ:Station investment} \begin{split} 
                IS_s = 1.3 * 600,000 EUR * \gamma * (\frac{C}{212kg/day})^{\alpha} \\ * (1-\beta)^{\log_2(\frac{C * n}{212kg/day*400})}
            \end{split}
            \end{equation} 
            
            \begin{table}
            \caption{Hydrogen fuel station assumptions}
            \label{tab:Hydrogen station assumptions}
            \begin{tabular}{llll}
            \toprule
            & LH2 & GH2 & LOHC \\
            \midrule
            {$\alpha$ $[-]$}  & 0.6 & 0.7 & 0.66  \\
            {$\beta$ $[-]$}  & 0.06 & 0.06 & 0.06  \\
            {$\gamma$ $[-]$}  & 0.9 & 0.6 & 1.4 \\
            EC $[kWh_{el}$/$kg_{H_2}]$ & 0.6 & 1.6 & 4.4\\
            NGC $[kWh_{NG}$/$kg_{H_2}]$ & 0 & 0 & 11.7 \\
            Depreciation years $[a]$ & 10 & 10 & 10\\
            O{\&}M [\%] & 5 & 5 & 5 \\
            \bottomrule
            \end{tabular}
            \end{table}
            
            Next, we identify the number and locations of fueling stations. Since passenger cars and trucks have different driving and refueling patterns, we separately select their fuel station locations.
            
            For passenger cars, we assume a utilization of 70{\%} and thus a turnover of 700 $kg_{H_2}$ per day, in line with \cite{Reuss2019}. This results in 412 fueling stations for cars. 
            We then first disaggregate the total demand to the >400 German NUTS-2 regions proportionally to the NUTS-2 gross domestic product (GDP). Since no more granular GDP data exists, we further break down the hydrogen demand to the over 10,000 NUTS-3 regions in Germany proportionally to the population in that NUTS-3 region. 
            Currently, there are 72 hydrogen fueling stations (October 2019) in Germany \citep{H2Live.2019}. Since these will not suffice to satisfy demand in 2030, we assume that additional fueling stations will be installed at the same locations as existing gasoline stations. Therefore, we use the 11,285 gasoline stations from OpenStreetMap as further potential sites \citep{OpenStreetMap}. For each of these stations, we calculate the distance to the closest NUTS-3 region center. For each NUTS-3 region, we then select stations with the shortest distance to its center, until its demand is covered.
            
            For trucks, \cite{Rose&Neumann2020} determine optimal hydrogen fuel station locations along highways under consideration of traffic flow and capacity limits. From these locations, we adopt those with highest utilization rate, which leads to 97 stations. We assume all fuel stations have 1,000 kg/day capacity and have the same turnover. Thus, to meet the demand from fuel cell heavy-duty trucks, the turnover of each fuel station is 847.42 $kg_{H_2}$ per day.
        
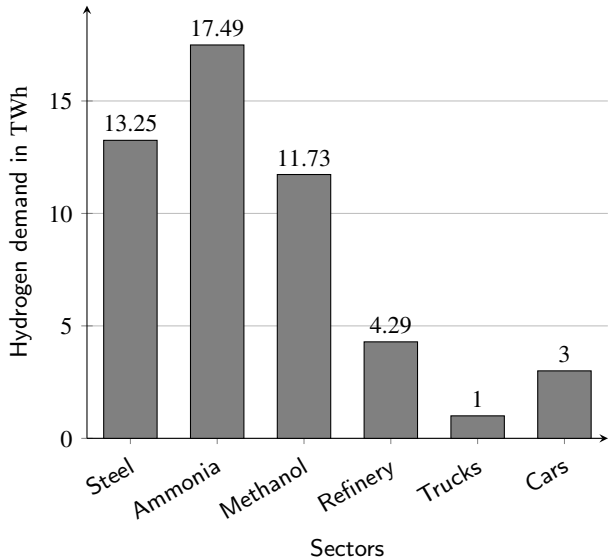
\begin{figure}[htb]
            \centering
            \begin{tikzpicture}
              \begin{axis}[
                ybar,
                bar width=20pt,
                nodes near coords,
                nodes near coords align=above,
                point meta=rawy,
                axis x line=bottom,
                axis y line=left,
                ymajorgrids=true,
                ylabel=Hydrogen demand in $\mathrm{TWh}$,
                ymin=0,
                ytick={0,5,10,15,20,25},
                enlargelimits=auto,
                xlabel= Sectors,
                symbolic x coords ={Steel,Ammonia,Methanol,Refinery, Trucks, Cars},
                x tick label style={rotate=30,anchor=north east},
                xtick distance=1,
                ]
            
                \addplot[fill=gray] coordinates {
                  (Ammonia,17.49)
                  (Steel,13.25)
                  (Methanol,11.73)
                  (Refinery,4.29)
                  (Trucks,1.00)
                  (Cars,3.00)
                };
              \end{axis} 
            \end{tikzpicture}
            \caption{Estimated hydrogen net demand per sector in Germany, 2030}
            \label{fig:hydrogen_demand_summary}
            \end{figure}

\begin{figure}[htbp]
            \centering
            \includegraphics[scale=.50]{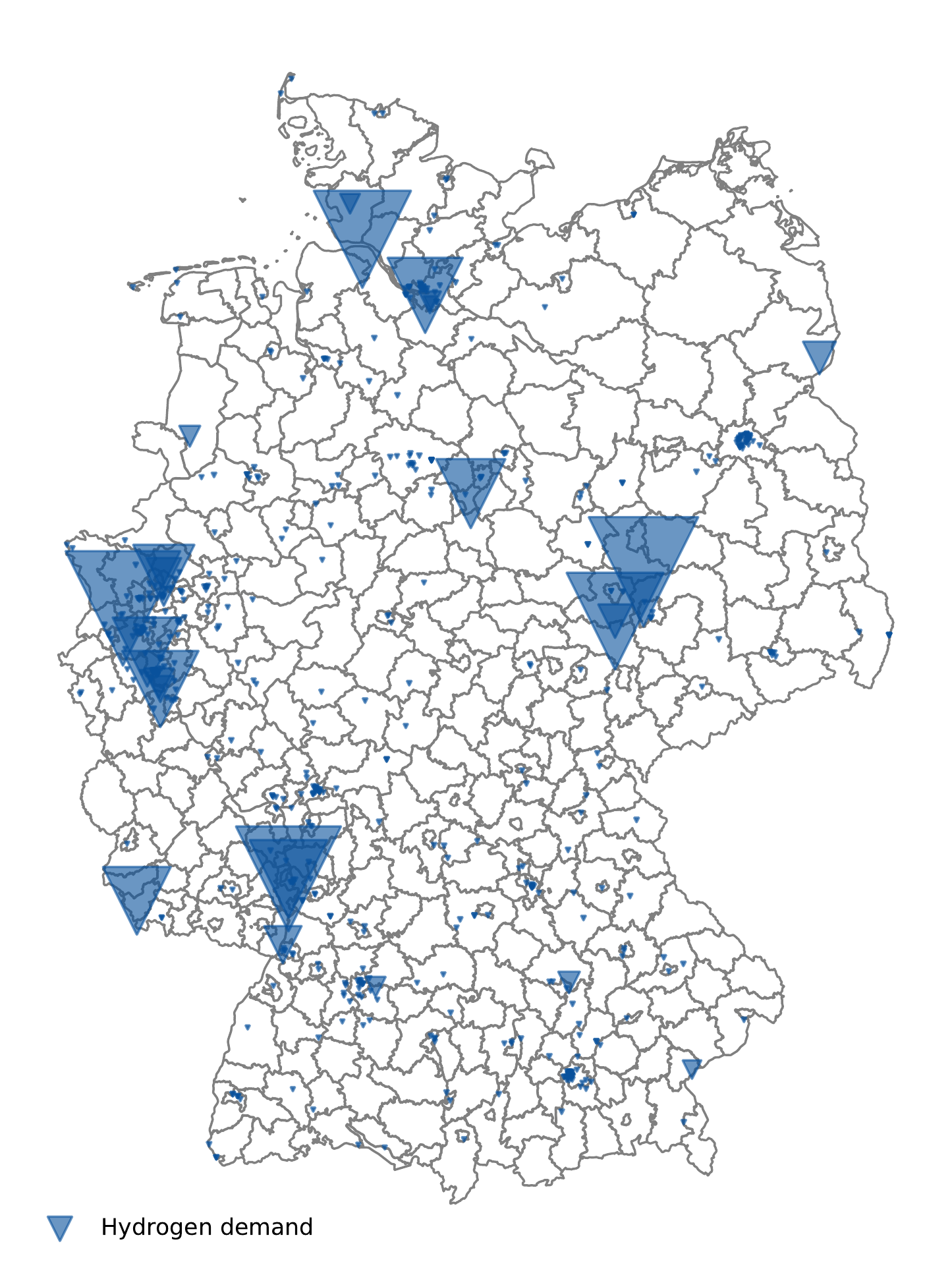}
            \caption{Spatial distribution of estimated hydrogen net demand in Germany, 2030}
            \label{fig:hydrogen_demand_map}
            \end{figure}

            \paragraph{\textbf{Summary of hydrogen demand}}        
            The total hydrogen net demand in 2030 is estimated to be 51.26 TWh. Figure \ref{fig:hydrogen_demand_summary} displays the hydrogen net demands of the individual sectors. The map in Figure \ref{fig:hydrogen_demand_map} shows the geographic distribution of the hydrogen demand, with the size of the markers corresponding to demand volume.\\
    
        \subsubsection{Hydrogen production and import data} \label{subsubsection:hydrogen_production_and_import_data}
            Electrolysis is the main pillar of political strategies for hydrogen supply in Germany \citep{Wasserstoffstrategie2020} and the European Union \citep{EU_hydrogen_strategy}. Among the different electrolysis technologies, proton exchange membrane (PEM) electrolysis is projected to have the lowest CAPEX and highest efficiency in 2030 \citep{bohm2020projecting}. Therefore, we focus on PEM electrolysis for hydrogen production. As input for the hydrogen supply chain model, we assume investment costs $IC_{Electrolyzer}$ of 604 EUR/$kW_{el}$, depreciation over 10 years, O{\&}M costs of 4\% of investment costs and electricity consumption $EC$ of 47.6 $kWh_{el}$ per $kg_{H2}$, based on \cite{Schmidt2017, Brown2018, Reuss2019}. Electrolyzer efficiency $EE$ is set to 70\% in line with \citep{heymann.2021, Robinius_2017_Part2, Reuss2019}. We set the minimum capacity $CAP_{Production, min}$ to 10 MW and the maximum capacity $CAP_{Production, max}$ to 100 MW. \footnote{This range was determined based on a review of power-to-gas projects that are scheduled to go be commissioned after 2021 in Germany (ELEMENT EINS, Energiepark Bad Lauchstädt, GET H2 Nukleus, HydroHub Fenne, GreenHydroChem Mitteldeutschland, Westküste100). All of them are in the range of 10-100 MW.} Regarding operation we analyze two different cases. In the main case, all electrolyzers are assumed to operate continuously under a flat tariff, at 70\% of full capacity, which is within typical ranges \citep{Robinius_2017_Part2, GUERRA20192425, Ruhnau2020Market}). In a sensitivity case, all electrolyzers are assumed to operate under a real-time tariff and have temporal flexibility which allows them to shift their operation to hours with cheap prices. In this case we assume they run at 100\% during the 70\% cheapest hours. Thus, in both cases, the total volume of produced hydrogen is the same.
            
            The potential locations for electrolyzers are equal to the set of transmission grid nodes from our electricity system model (compare Section \ref{subsub:transmission_grid_data}).\footnote{Such large-scale electrolyzers might be complemented by smaller, on-site electrolyzers (see, e.g. \citet{Rose&Neumann2020, Golla2020}) in practice. Such on-site electrolyzers would be connected to the distribution grid. Analyzing congestion consequences at distribution grid level is out of scope of this study.} Besides, we include hydrogen imports from overseas into our model, since they are a key part of the German hydrogen strategy \citep{Wasserstoffstrategie2020}. For these imports, we assume a fixed, exogenous amount of daily available imported hydrogen of 27.40 GWh and costs of 3.48 EUR/$kg_{H_2}$, in line with the mean values reported by \cite{runge2020}. Furthermore, we assume that all imports to Germany will occur at one large port, i.e. Bremerhaven, in line with \cite{runge2020}.
        
        \subsubsection{Hydrogen conversion data}
            Hydrogen can be converted to a liquefied state (LH2), compressed state (GH2), or stored into chemicals (LOHC) for transportation via tube trailers. Notably, for LH2 and LOHC, there are capital and operating costs at the point of hydrogen production (for liquefaction, and hydrogenation, respectively) and at the point of hydrogen consumption (evaporation, and dehydrogenation, respectively).
            
            The assumptions regarding investment costs, depreciation years, O\&M costs, electricity and natural gas consumption, and losses are displayed in Table \ref{tab:Conversion assumptions}.
            
            \begin{table*}
                \caption{Conversion assumptions, based on \cite{Reuss2019, Nexant.2008}. $x$ denotes daily hydrogen output.}
                \label{tab:Conversion assumptions}
                \centering
                \begin{tabular}{L{2.2cm}L{4.2cm}L{1.7cm}L{0.7cm}L{2cm}L{2cm}L{1.2cm}}
                \toprule
                & Investment costs [EUR] & Depreciation years & O{\&}M & $EC_{Conversion}$ $[kWh_{el}/kg_{H_2}]$ & $NGC_{Conversion}$ $[kWh_{NG}/kg_{H_2}]$ & Loss $[\%]$ \\
                \midrule
                Compressor & $15 * 10^3 \frac{EUR}{kW} * x^{0.6089} * 3 $ & 15 & 4{\%} & calculated & 0 & 0.5 \\
                Liquefaction & $105 * 10^6 EUR * (\frac{x}{50\frac{t_{H_2}}{day}})^{0.66}$ & 20 & 4{\%} & 6.78 & 0 & 1.65\\
                Evaporation & $3 * 10^3 EUR * \frac{x}{1,000}$ & 10 & 3{\%} & 0.6 & 0 & 0\\
                Hydrogenation & $40 * 10^6 EUR * (\frac{x}{300\frac{t_{H_2}}{day}})^{0.66}$ & 20 & 3{\%} & 0.37 & 0 & 1 \\
                Dehydrogenation & $30 * 10^6 EUR * (\frac{x}{300\frac{t_{H_2}}{day}})^{0.66}$ & 20 & 3{\%} & 0.37 & 11.7 & 1 \\
                \bottomrule
                \end{tabular}
            \end{table*}
        
        \subsubsection{Hydrogen transportation data}
            Transportation costs include costs for fuel, toll, and the drivers' wages. We assume that delivery trucks are fueled with hydrogen. The consumption is set to 5.19 $kg_{H_2}$/100km and the fuel price to 7,91 EUR/kg, including value added tax of 19\% \citep{BZ-LKW2017}. In line with \cite{Reuss2019Dissertation}, we make the following cost assumptions. Toll is set to 0.15 EUR/km. Drivers' wage is set to 35 EUR/h. Average driving speed is set to 50 km/h. Truck investment costs are set to 174,000 EUR \citep{BZ-LKW2017}, with depreciation over eight years and 12\% O\&M costs. For tube trailers, investment costs and capacities are technology specific. They are set to 860,000 EUR and 4,300 $kg_{H_2}$ for liquefied hydrogen (LH2), 660,000 EUR and 1,100 $kg_{H_2}$ for gaseous hydrogen (GH2), and to 150,000 EUR and 1,620 $kg_{H_2}$ for LOHC. Besides, we assume depreciation over twelve years and O\&M costs of 2\%, adopted from \cite{Reuss2019}.
    
        \subsection{Electricity system data} \label{subsub:electricity_data}
            We parametrize both, the uniform price and the nodal price electricity model with data for generation, consumption and the transmission grid in 2030. For this, we utilize the data set published by \citet{data2030}. In the following, we briefly describe this data set. All data are more elaborately documented and available for free use under a Creative Commons license in \cite{data2030}.

            \subsubsection{Transmission grid data} \label{subsub:transmission_grid_data}
                The transmission grid in 2030 is constructed from the reference ELMOD model of the existing grid, which is enhanced with all the expansions and new installations until 2030 that have been announced by the German Federal Network Agency. The resulting final grid representation consists of 485 nodes and 663 lines. The transmission capacity of all 220 kV lines is set to 490 MW, and that of all 380 kV lines to 1700 MW, based on \cite{Egerer2016Open, kiessling2011freileitungen}. 
            
            \subsubsection{Electricity demand data}
                For consumption, the hourly consumption forecast scenario EUCO30 is used \citep{EuropeanNetworkofTransmissionSystemOperatorsforElectricity.2018}. To improve consistency of grid and consumption data, these hourly values are re-scaled so that the annual total (577 TWh) matches the sum used in the official grid development plan (544 TWh) by \cite{NEPBest.2019}.\\
                
                Next, these re-scaled hourly demand values are spatially disaggregated to NUTS-3 levels. For this  disaggregation, the gross domestic product (GDP) and the population of a region serve as proxies for its future electricity consumption. The resulting NUTS-3 consumption time series are assigned to the nearest grid node.
            
            \subsubsection{Electricity generation data}
                For generation, estimation is differentiated between renewable, i.e. non-dispatchable generation, and dispatchable generation.\\
                
                For renewable generation, i.e. solar and wind, historical hourly generation data from the four national grid operators \citep{Bundesnetzagentur.2018} is used. For the baseline scenario with no hydrogen, these hourly values are re-scaled so that the annual total generation from each technology matches the sum used in the grid development plan \citep{NEPBest.2019}. This results in an annual generation of 86.7 TWh from solar (compared to an mean of 35.34 TWh in 2016-2018), and of 247.4 TWh from wind (compared to an mean of 108.6 TWh in 2016-2018). For the scenarios with hydrogen we factor in the current discussion about ''additionality'' \citep{Pototschnig.2021}, by further scaling up the capacity of solar and wind proportionally to the additional electricity demand for hydrogen production. Last, the re-scaled hourly generation values are spatially disaggregated. For this, we use the installed generation capacity per zip code as provided by \cite{UNB.2018}.\\
                
                For dispatchable electricity generation capacity, all relevant plants for 2030 from the power plant list of the German grid regulator are used \citep{BNetzA-power-plant-list}. For each plant, marginal costs are calculated, based on fuel type, estimated efficiency, and emissions costs. A $CO_2$ price of 60 EUR/ton is assumed \citep{Bundesregierung.2019}.\\
                
                Both renewable generation time series and dispatchable power plants, along with their marginal costs, are assigned to the nearest grid node. Note that this approach provides high spatial granularity, but comes at the costs of treating Germany as an isolated system without cross-border electricity lines. This can affect the results for electricity prices and redispatch in both directions, as noted by \cite{XIONG2021116201}. Therefore, a geographic expansion -- e.g. a European model -- can be worthwhile future work, but requires substantial additional data procurement efforts if the high spatial granularity (i.e. 485 nodes in the German system) is to be upheld.\footnote{A starting point could be the open network model PyPSA-Eur-Sec-30 that works with one node per country \citep{VICTORIA2019111977}.}

\section{Results and Discussion}
    Upon parametrizing the models presented in Chapter \ref{Section:Methodology} with the case study data presented in Chapter \ref{Section:Case_Study}, we run the models sequentially in three steps. First, we derive baseline results for the electricity system without hydrogen, including wholesale uniform prices, nodal prices and congestion management costs (Chapter \ref{subsection:electricity_results}). Second, based on the resulting electricity tariffs, we derive information about the optimal hydrogen supply chains, including total end-use costs of hydrogen, as well as number, capacities and locations of electrolyzers (Chapter \ref{subsection:hydrogen_results}). Third, we observe the effects of integrating these hydrogen supply chains in the electricity system, including changes in total electricity demand, wholesale prices, and redispatch costs (Chapter \ref{subsection:integration_results}).
    
    \subsection{Baseline electricity system results}\label{subsection:electricity_results}
\begin{figure}[htbp]
        \centering
        \includegraphics[scale=.25]{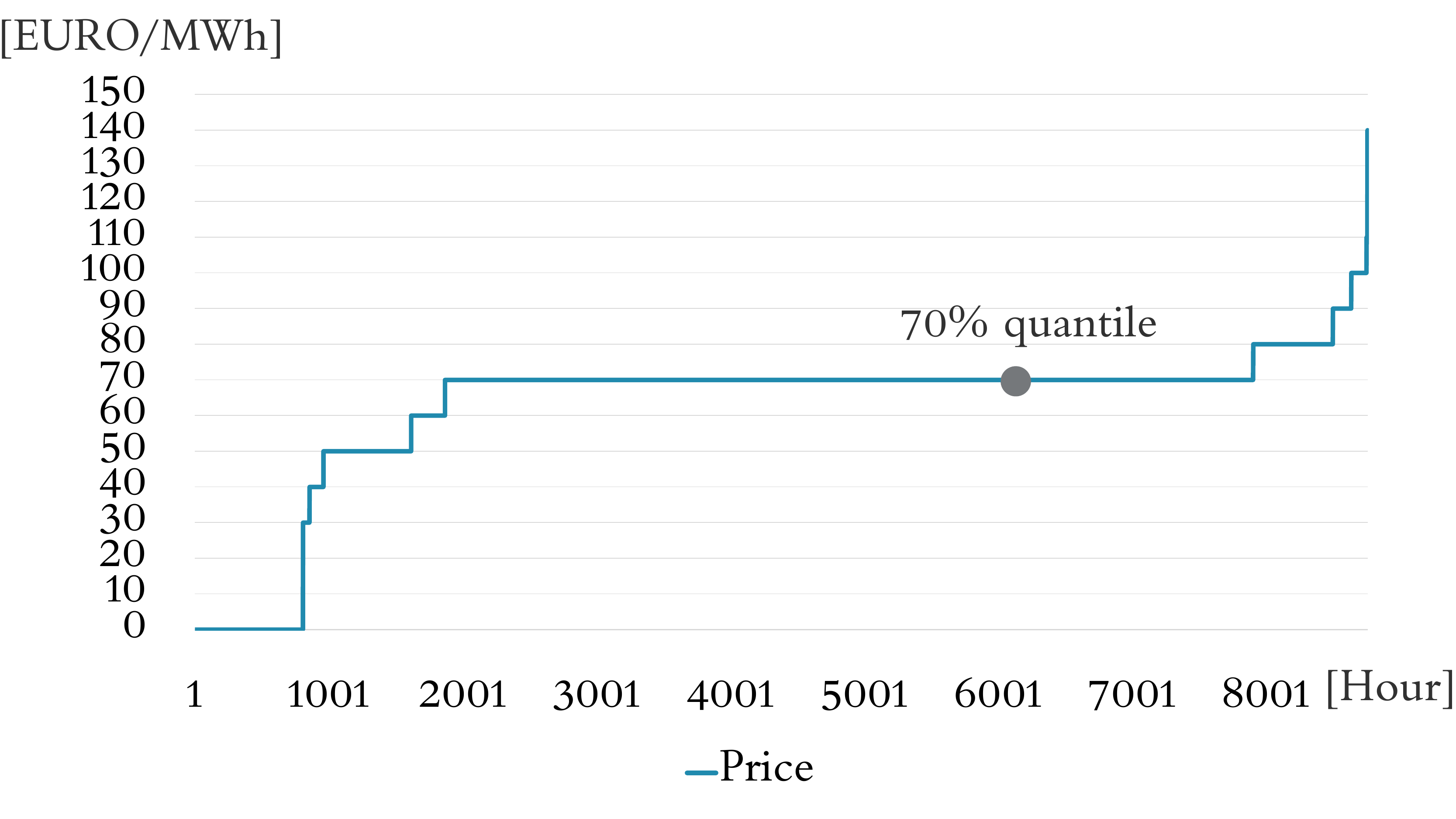}
        \caption{Wholesale price duration curve in Germany, 2030 [EUR/MWh]}
        \label{fig:wholesale_prices}
        \end{figure}

        Without the integration of hydrogen the resulting annual mean of the wholesale uniform price is 62.61 EUR/MWh. Figure \ref{fig:wholesale_prices} shows the price duration curve. The annual redispatch costs amount to 6.16 Billion EUR.
        
\begin{figure}[htbp]
        \centering
        \includegraphics[scale=.50]{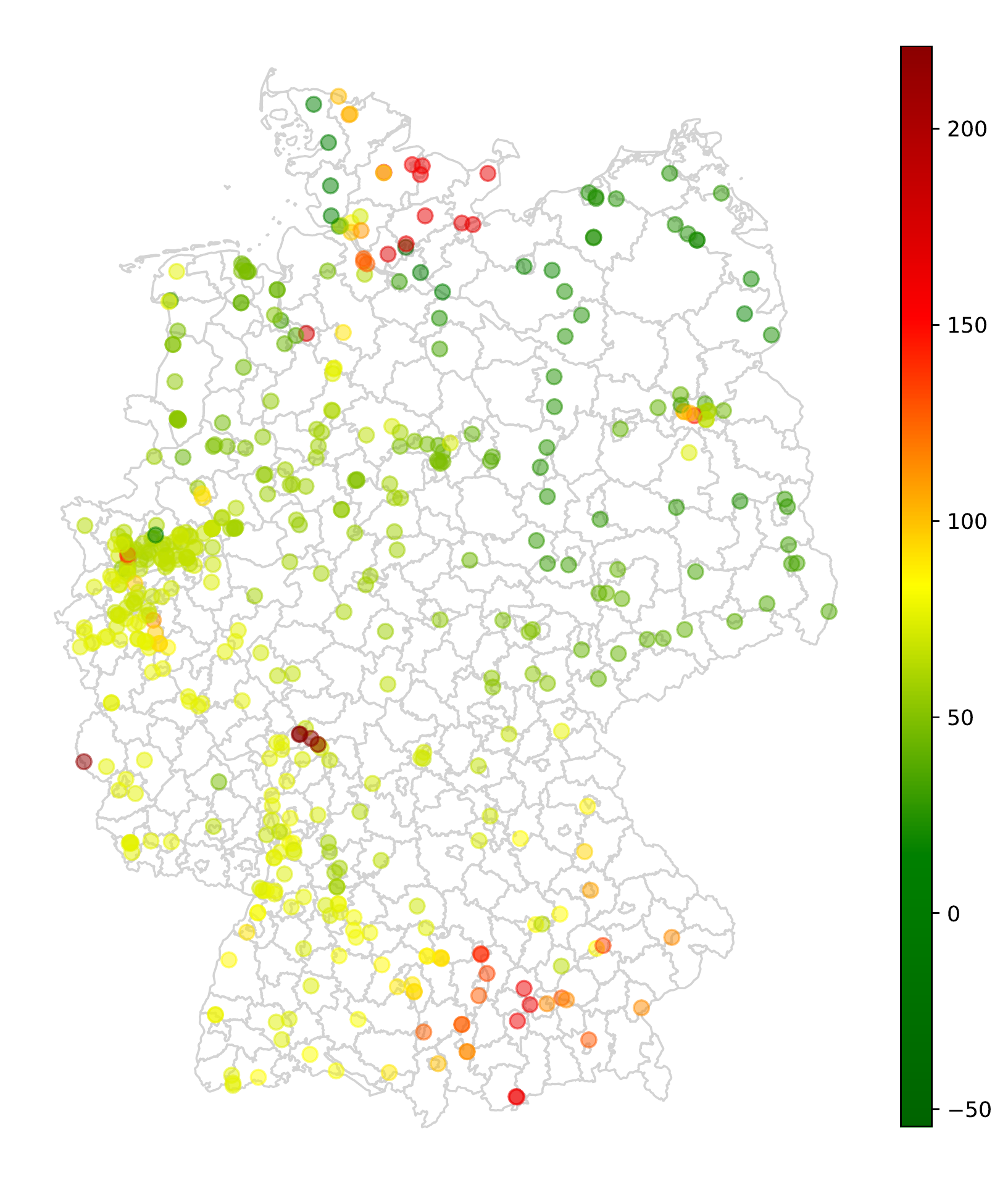}
        \caption{Shadow nodal prices in Germany, 2030 [EUR/MWh]}
        \label{fig:nodal_prices}
        \end{figure}

        The resulting annual means of nodal prices vary between -54.30 and +221.00 EUR/MWh, with a median value of 67.80 EUR/MWh. Figure \ref{fig:nodal_prices} shows the spatial distribution of nodal prices. Low prices are predominantly found in the North-East and North-West of the country, driven by high renewable generation and low demand. This finding is in line with \citet{Robinius_2017_Part2} who analyze residual loads on county level and find negative residual loads predominantly in the North-East and North-West.\footnote{\cite{NEUHOFF2013760} calculate nodal prices for the year 2008 and find that prices vary between 10 and 100 €/MWh. The geographic disparity in the German generation system strongly increases from 2008 to 2030 due to a complete shut-down of all nuclear power plants, the partial shut-down of coal power plants, a strong renewable expansion especially in the North, and the introduction of a carbon emission price. This presumably causes the prices in our 2030 case study to have a larger (geographic) variance.}
        
    \subsection{Hydrogen supply chain results} \label{subsection:hydrogen_results}
\begin{figure*}
        \centering
        \includegraphics[scale=.60]{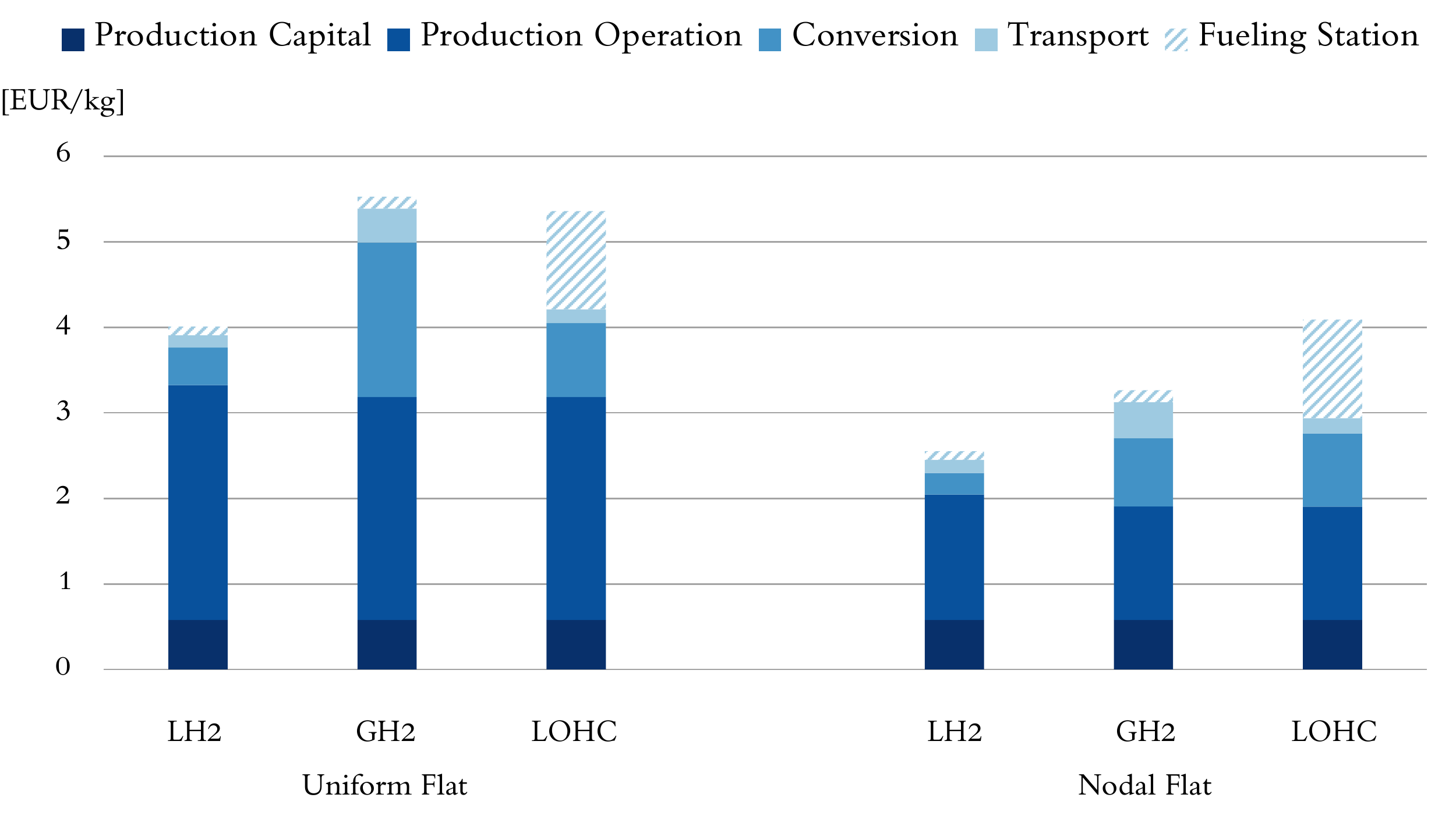}
        \caption{End-use hydrogen costs by component and scenario: Effects of delivery form}
        \label{fig:cost_components}
        \end{figure*}

        The resulting end-use costs for hydrogen are represented in Figure \ref{fig:cost_components}. As noted above, it is assumed that the tariff equals the energy price, omitting grid fees, taxes, and other charges. In the uniform flat scenario, which implies static electrolyzer operation under a time-invariant price, the final hydrogen costs for industry applications are 3.91 $EUR/kg_{H_2}$ for LH2, 5.39 $EUR/kg_{H_2}$ for GH2, and 4.21 $EUR/kg_{H_2}$ for LOHC. For fuel cell trucks and cars, costs for fueling stations need to be added, resulting in final hydrogen costs of 4.01 $EUR/kg_{H_2}$ for LH2, 5.53 $EUR/kg_{H_2}$ for GH2, and 5.36 $EUR/kg_{H_2}$ for LOHC. The largest share of costs are caused by production operation in all cases. These are largely driven by electricity costs (compare Eq. \ref{equ:POC}), which highlights the large role of electricity prices for the end-use costs of electrolytic hydrogen.
        
        This is also reflected by the results under the nodal flat tariff. In this case, the lower electricity costs for electrolyzers lead to much lower total hydrogen costs, i.e. to 2.45 (2.55 for fuel cell trucks and cars) EUR/$kg_{H_2}$ for LH2, 3.13 (3.26) EUR/$kg_{H_2}$ for GH2, and 2.94 (4.09) EUR/$kg_{H_2}$ for LOHC.\\

        For the cheapest delivery form, i.e. LH2, we additionally compute the scenarios with flexible operation under real-time tariffs. In these scenarios electrolysis is shifted to hours with the lowest prices. For the uniform real-time case, electrolyzers are assumed to run at 100\% capacity during the 70\% cheapest hours at the wholesale market. For the nodal real-time case, electrolyzers are similarly assumed to run at 100\% capacity during the 70\% cheapest hours of the respective node.
        
\begin{figure*}
        \centering
        \includegraphics[scale=.60]{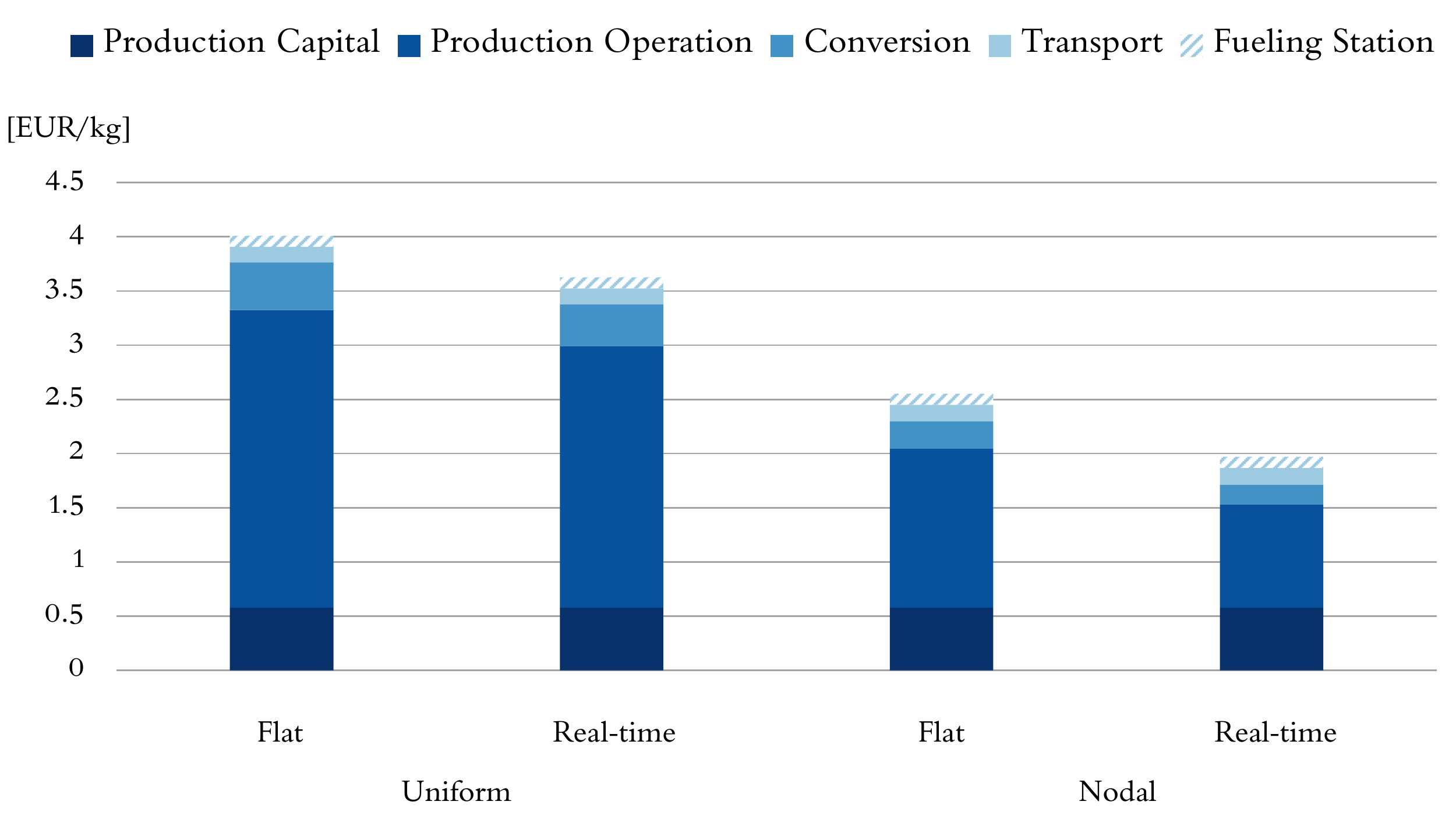}
        \caption{End-use hydrogen costs by component and scenario: Effects of flat versus real-time tariffs}
        \label{fig:cost_components_flat_v_realtime}
        \end{figure*}

        This flexible operation enables electrolyzers to use cheaper electricity and thus leads to overall lower hydrogen costs, as Figure \ref{fig:cost_components_flat_v_realtime} shows. Moving from uniform flat to uniform real-time tariffs decreases total costs by 0.38 $EUR/kg_{H_2}$. Moving from static operation under the nodal flat tariff to flexible operation under the nodal real-time tariff decreases costs by 0.58 $EUR/kg_{H_2}$.\\

\begin{figure*}[htbp]
              \begin{subfigure}[c]{0.3\textwidth}
              \includegraphics[height=7cm]{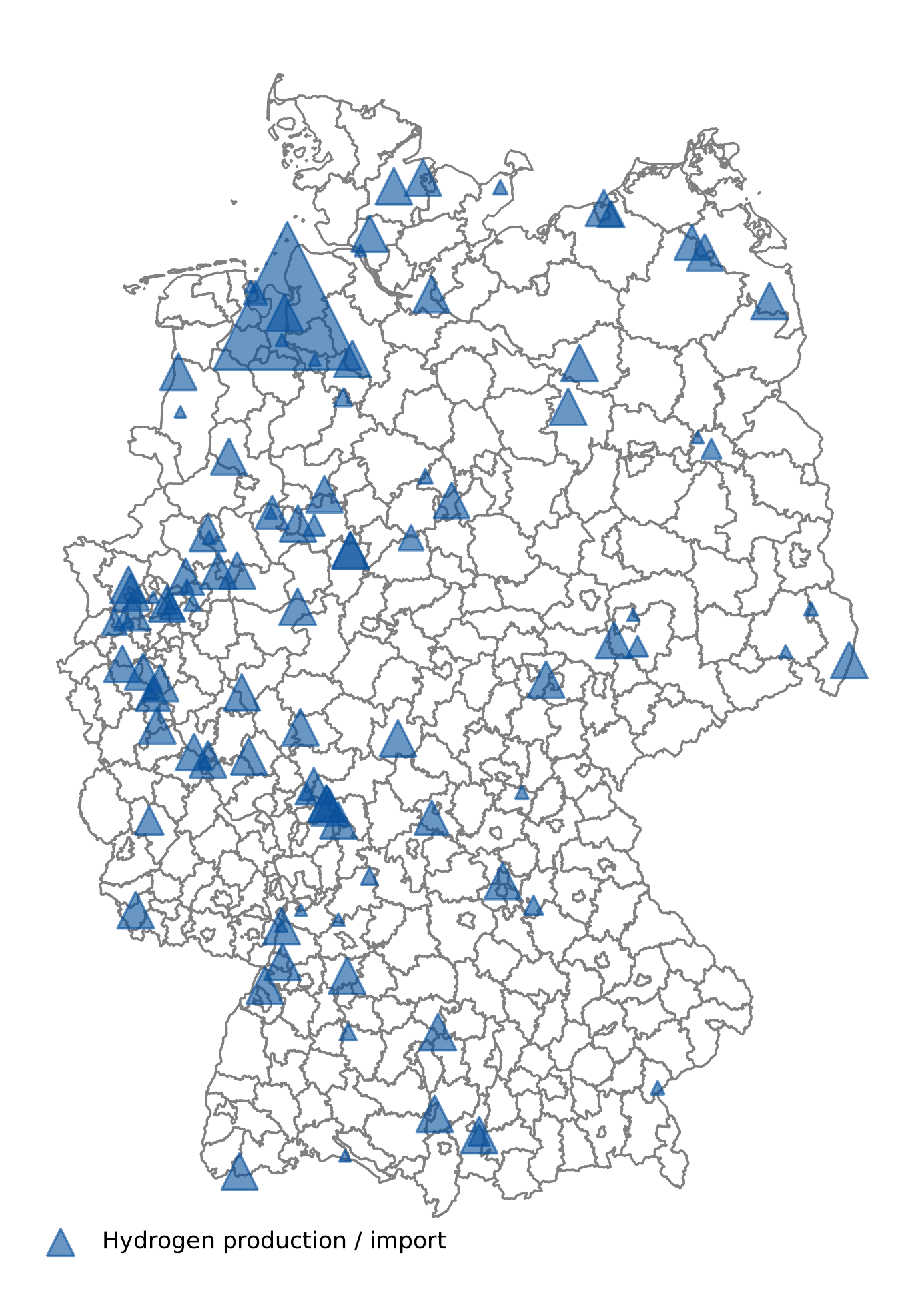}
              \caption{Delivery with LH2 Trailers}
              \label{fig:LH2_uniform}
              \end{subfigure}
              \begin{subfigure}[l]{0.3\textwidth}
              \includegraphics[height=7cm]{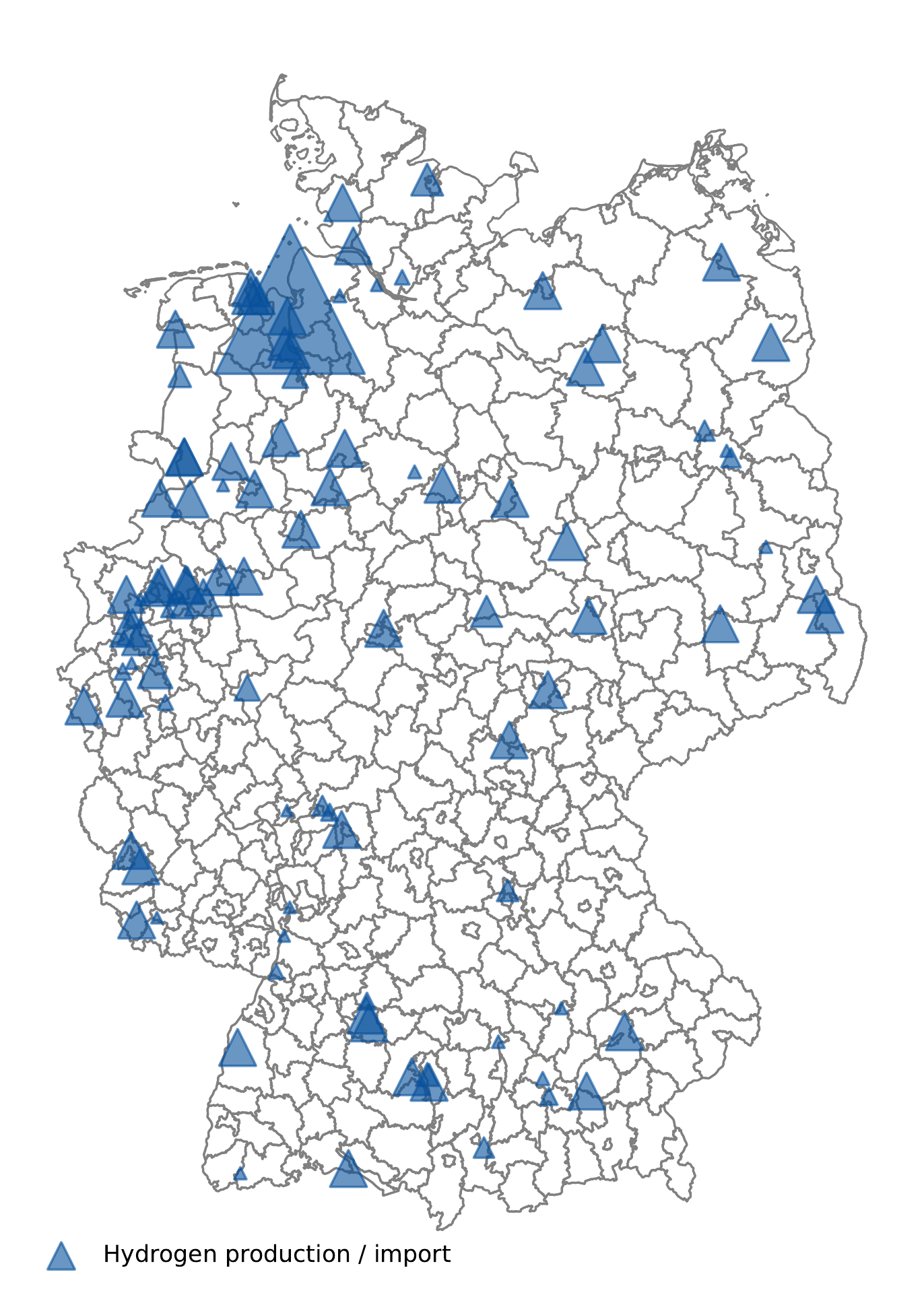}
              \caption{Delivery with GH2 Trailers}
              \label{fig:GH2_uniform}
              \end{subfigure}
              \begin{subfigure}[r]{0.3\textwidth}
              \includegraphics[height=7cm]{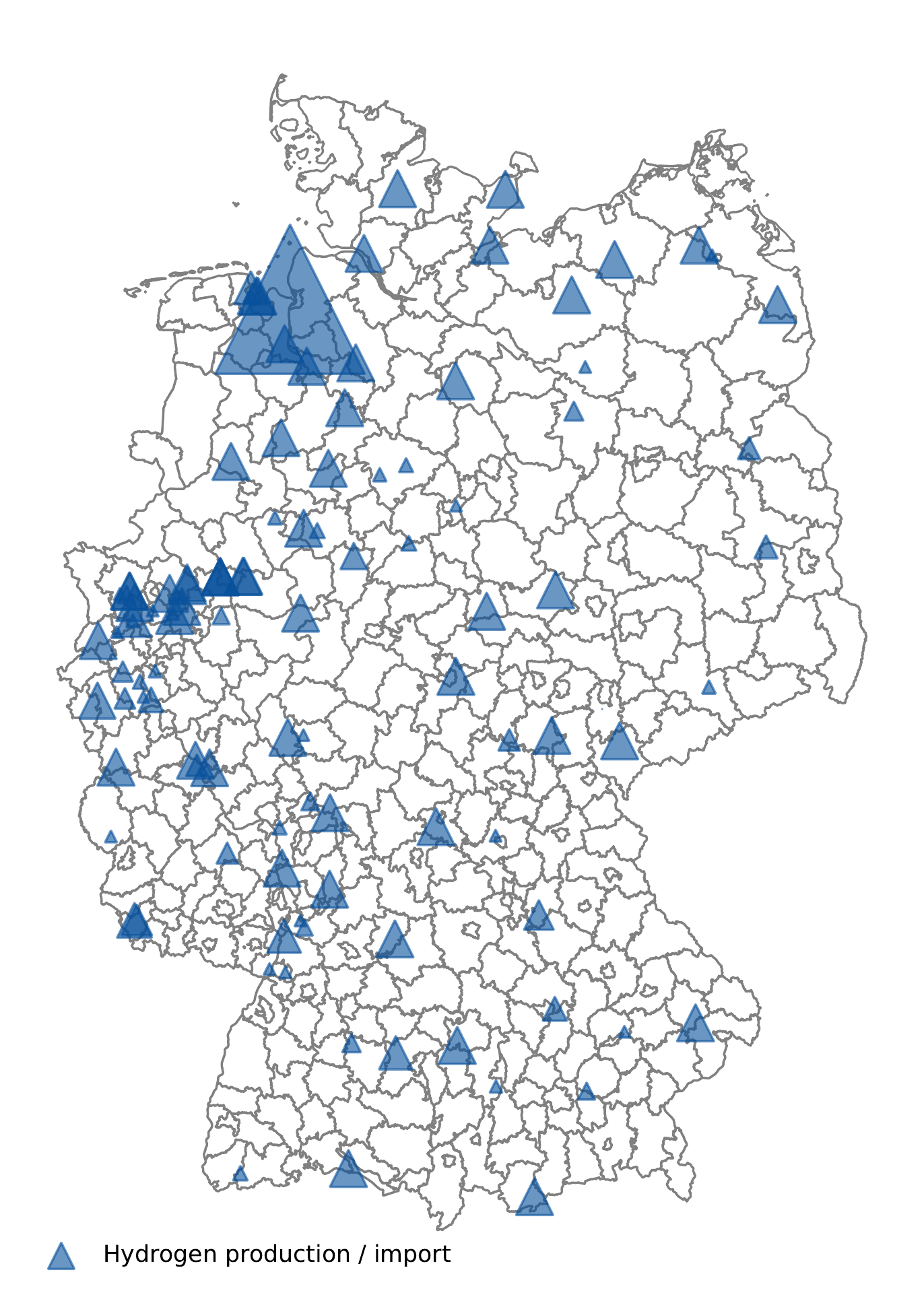}
              \caption{Delivery with LOHC Trailers}
              \label{fig:LOHC_uniform}
              \end{subfigure}
        \caption{Optimal electrolyzer locations under Uniform Flat tariff}
        \label{fig:optimal_production_uniform}
        \end{figure*}

\begin{figure*}[htbp]
              \begin{subfigure}[l]{0.3\textwidth}
              \includegraphics[height=7cm]{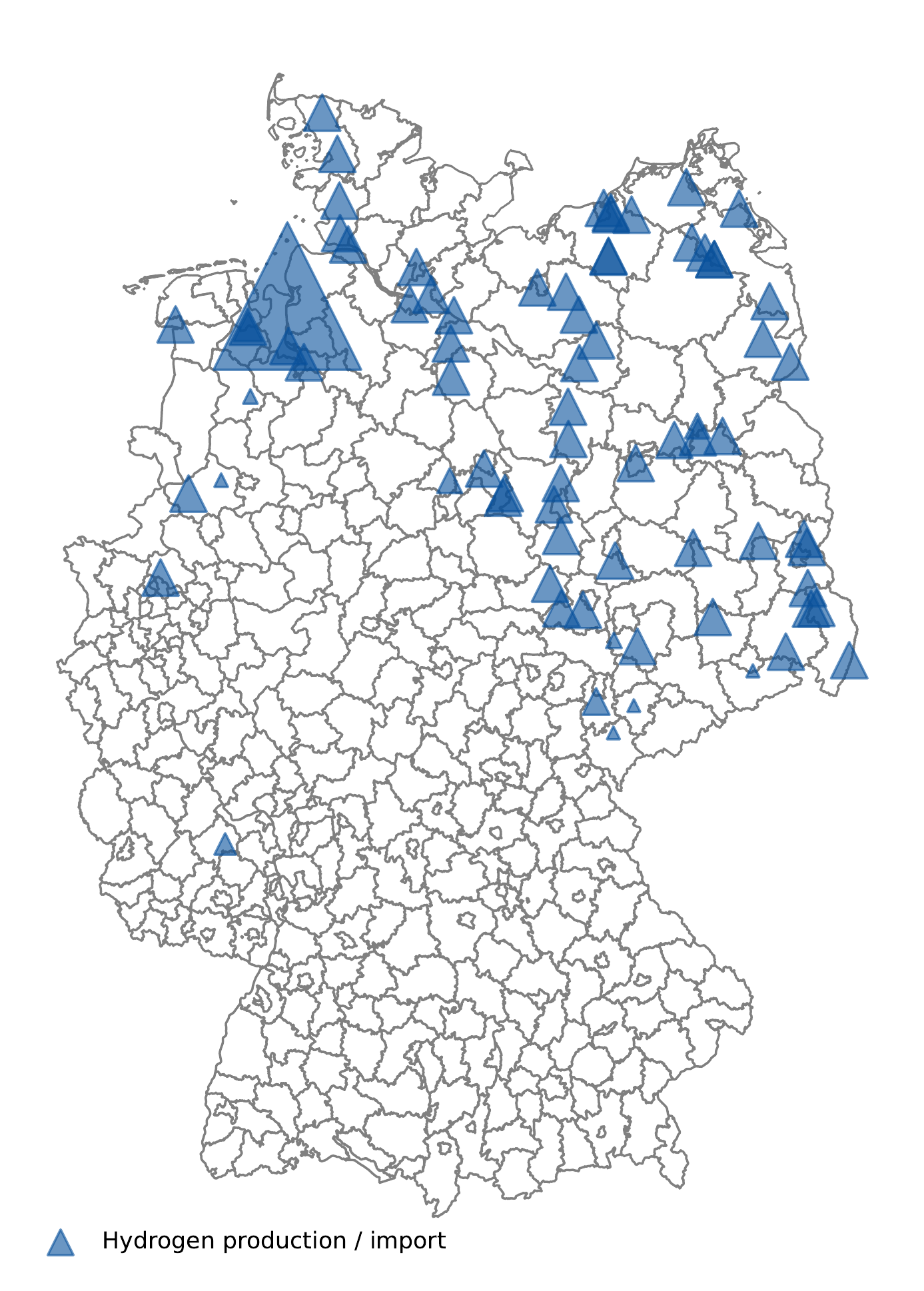}
              \caption{Delivery with LH2 Trailers}
              \label{fig:LH2_nodal}
              \end{subfigure}
              \begin{subfigure}[c]{0.3\textwidth}
              \includegraphics[height=7cm]{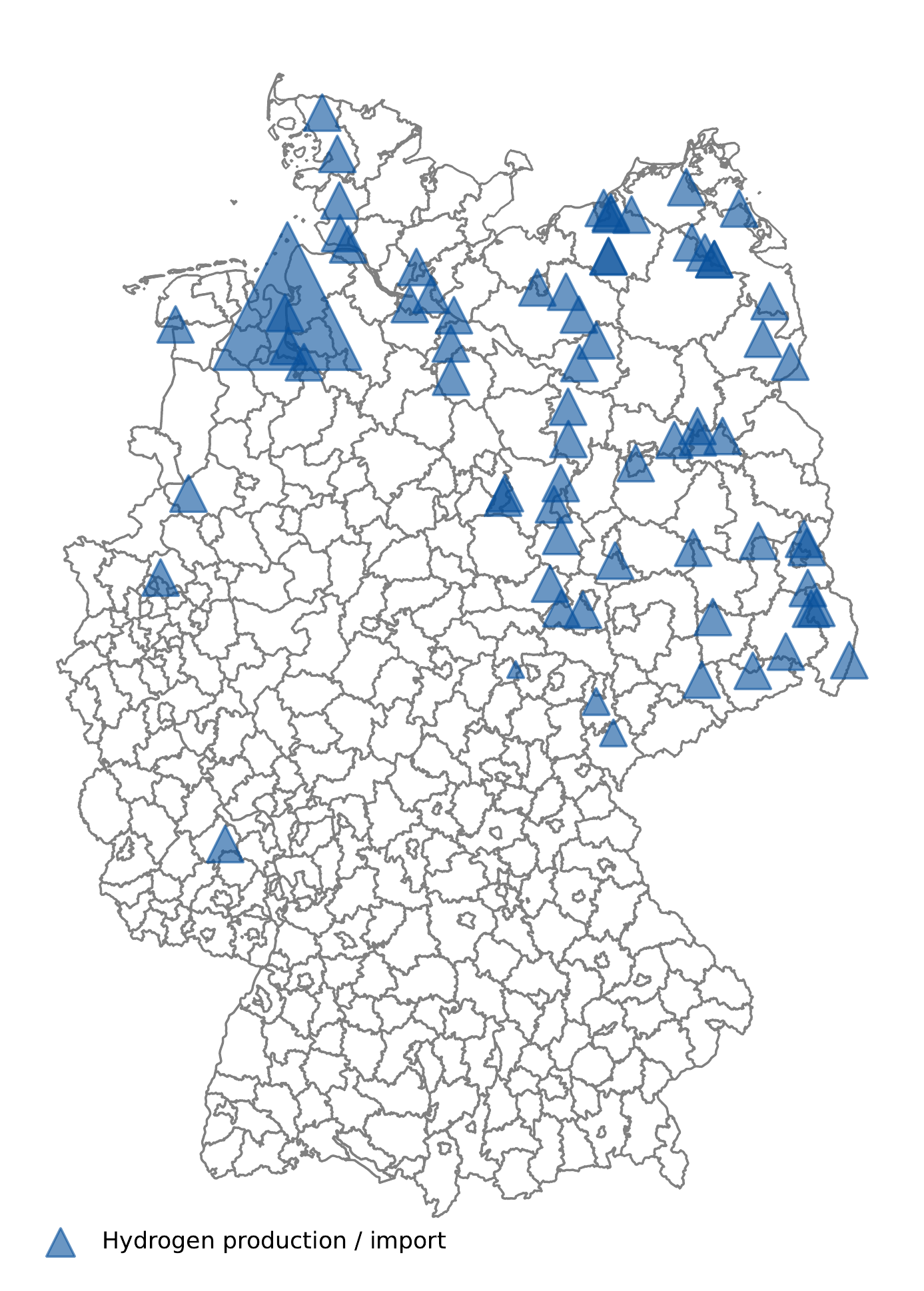}
              \caption{Delivery with GH2 Trailers}
              \label{fig:GH2_nodal}
              \end{subfigure}
              \begin{subfigure}[r]{0.3\textwidth}
              \includegraphics[height=7cm]{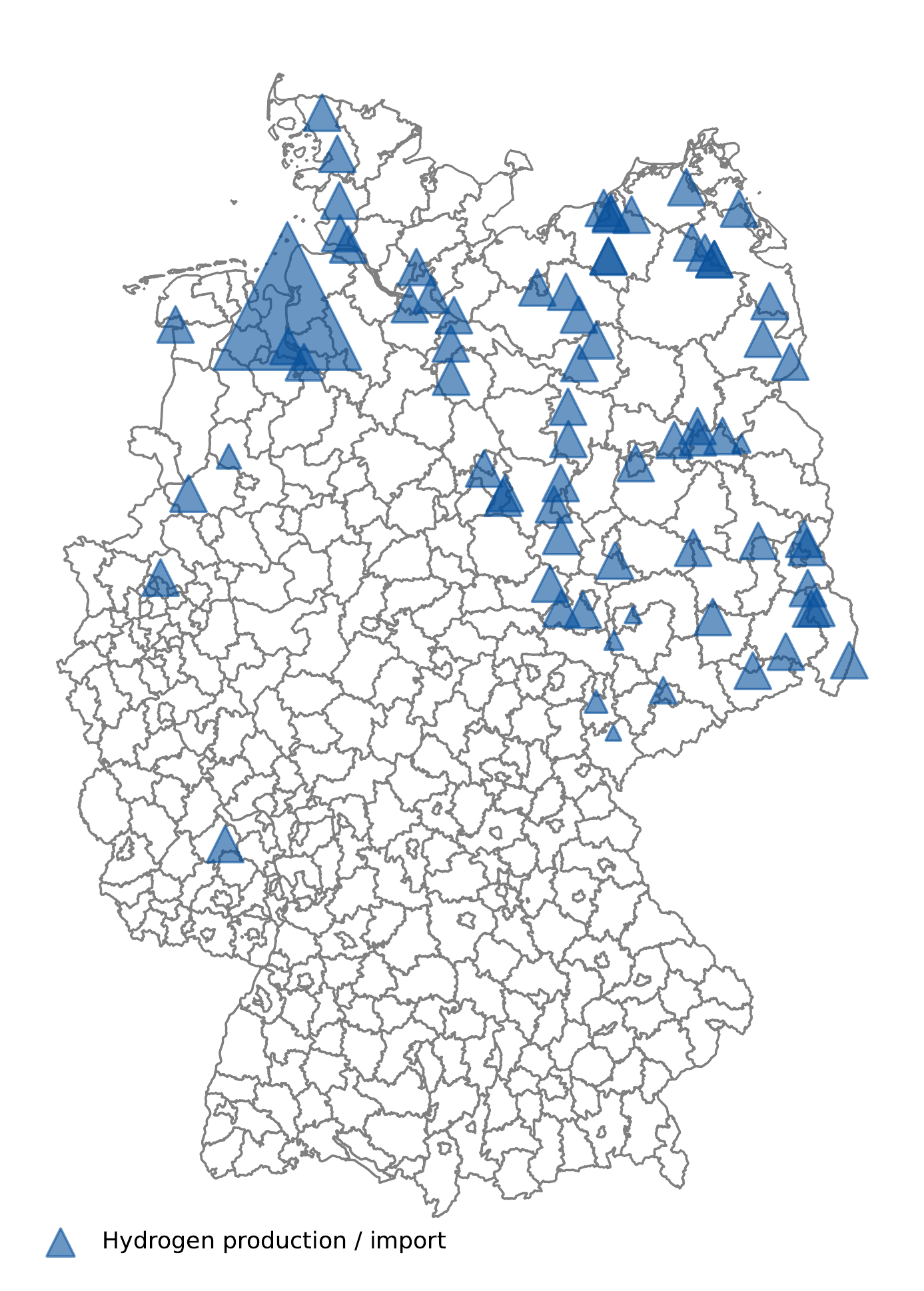}
              \caption{Delivery with LOHC Trailers}
              \label{fig:LOHC_nodal}
              \end{subfigure}
        \caption{Optimal electrolyzer locations under Nodal Flat tariff}
        \label{fig:optimal_production_nodal}
        \end{figure*}

        The cost-minimal locations of electrolyzers are depicted in Figures \ref{fig:optimal_production_uniform} and \ref{fig:optimal_production_nodal}. The size of the markers corresponds to production volume. The largest marker in the North-West depicts overseas imports, which are exogenously determined (compare Chapter \ref{subsubsection:hydrogen_production_and_import_data}) and thus occur equally in all scenarios. In terms of domestic production, 9.50 $GW_{el}$ of electrolysis capacity are installed in all scenarios with demand from industry, trucks, and cars.
        
        Under uniform flat tariffs with LH2, 101 domestic electrolyzers are placed. They are predominantly placed close to points of consumption, in order to minimize transportation costs (Figure \ref{fig:optimal_production_uniform}). Half of the installed electrolyzers (52) have the maximal possible capacity of 100 MW. The siting is very similar for GH2 and LOHC, with 97, and 106 electrolyzers placed, respectively.
        
        Under nodal flat tariffs, electrolyzers are placed further away from consumption, but at nodes with low electricity prices (Figure \ref{fig:optimal_production_nodal}). This indicates that the cheaper electricity costs outweigh the higher transportation operating costs. This effect is stable across the three delivery states, and is in line with the findings from \cite{vomscheidt.2021, Robinius_2017_Part2, JENTSCH2014254}. In the LH2 nodal case, 76 electrolyzers are placed, of which 63 have maximal possible capacity. Again, the siting is very similar for GH2 (69 electrolyzers), and LOHC (72).
        
        In both scenarios there are small differences between the three delivery states, which are due to the different trailer capacities, different per-kg transport operating costs (see Eq. \ref{equ:TOC} to \ref{equ:FTC}) and conversion operating costs (see Eq. \ref{equ:COC_LH2} to \ref{equ:COC_GH2}). These in turn affect the optimal location of electrolyzers (compare Eq. \ref{equ:total annual cost}). 
            
    \subsection{Integration results}\label{subsection:integration_results}
        From the above presented locations and capacities of electrolyzers, we calculate the additional electricity demand from hydrogen production at each grid node. With this new input, we recalculate electricity prices and congestion management costs to identify the effects of hydrogen on the electricity system. For these calculations we assume LH2 delivery, since it is the cost-minimal hydrogen supply chain set-up in all scenarios, for both industry and transportation applications.
        
        Table \ref{tab:feedback_results} summarizes the key results. The electrolytic production of hydrogen creates considerable new electricity demand of 72.49 TWh per year that increases the total national electricity demand by about 13\%. Since this demand is assumed to be met by additionally installed solar and wind capacity, the average wholesale prices slightly decrease, by 4-7\%, depending on the scenario. 
        
        If the hydrogen supply chain is optimized according to the uniform tariff, annual congestion management costs \textit{increase} by 17-18\%. This corresponds to an increase of over one billion Euros per year. Interestingly, this increase is only slightly smaller for the uniform real-time tariff with flexible operation, compared to the uniform flat tariff with static operation. This finding indicates that electrolyzers which respond to real-time wholesale prices, but are inefficiently placed from a system perspective, might not be able to fully realize expected positive impacts regarding the actual use of cheap renewable energy (cf. e.g. \citep{Ruhnau2020Market}) due to grid constraints. A key explanatory factor for this might be that wind generation, which to a large extend is located in the North of Germany \citep{UNB.2018} has been shown to drive wholesale prices down \citep{Benhmad_2018} and at the same time drives congestion in the transmission grid \citep{Staudt_2019}. This finding motivates additional in-depth analyses of the interplay of spatial and temporal dimension for successful hydrogen integration in the electricity system.
        
        In contrast, when electrolyzers are placed and operated under nodal price signals, they \textit{decrease} congestion management costs by 17-20\%. This decrease corresponds to over one billion Euro per year. The decrease is slightly larger under the nodal real-time tariff which shows that the combination of resolved spatial and temporal signals yields largest benefits.
        
        In summary, there is a welfare delta of over two billion Euros per year between hydrogen integration under the status quo uniform price scenarios and under the nodal price scenarios. In other words, the production of one $kg_{H_2}$ on average creates additional congestion costs of 0.68-0.72 Euros under current regulation, whereas it reduces congestion costs by up to 0.82 Euros under more efficient regulation. This means a spatially differentiated subsidy for hydrogen production -- e.g. in the form of a per-kWh payment of the spread between uniform prices and simulated nodal prices -- could effectively be covered by saved redispatch costs.
        
        It is noteworthy, that the model does not assume that investors consider how their electrolyzer installation will affect nodal prices. This would require iterative calculation of both models, which is out of scope due to high computational effort.
        Regarding practical implementability, the complex and potentially vulnerable nature of nodal price signals represent another limitation. For hydrogen investors to base their decisions on nodal price signals, they need to be able to forecast these a priori, for which sufficient information and appropriate data analytics methods need to be available \citep{vomScheidt_2020_Review}. Furthermore, investors face the risk of unforeseen expansion of grid or generation capacity that impacts nodal prices. Therefore, policy makers could opt to assess less efficient price signals such as zonal tariffs and regionalized grid fees. Alternatively, non-price mechanisms are conceivable, such as regional quotas or allowing grid operators to curtail electrolyzers before performing the regular redispatch measures, in case of congestion. While such mechanisms typically forego some of the efficiency gains from nodal signals, they provide advantages regarding simplicity, and associated risks. Therefore, it is desirable that future studies investigate the advantages and disadvantages of different (spatial) regulatory instruments.\footnote{For related analyzes regarding the integration of generation see e.g. \cite{Bertsch2015Congestion, grimm2019regionally, Schmidt2020price}. For a comprehensive review of locational investment signals for generation capacity that are applied in practice see \cite{eicke2020locational}.}
        
        Another interesting expansion of this work lies in quantifying the effects of electrolysis on system-wide greenhouse gas emissions. While hydrogen from electrolysis will decrease emissions in the end-usage sectors steel, ammonia, methanol, refineries, and transportation, it is relevant to quantify the emission effects of hydrogen production on the electricity system. We expect that the decrease in congestion enabled by electrolyzers placed according to nodal tariffs will also lead to a decrease in system emissions, since a considerable, and rising share of curtailed energy comes from renewable sources \citep{XIONG2021116201} and a reduction of redispatch in German regions with high renewable shares leads to overall emission reductions \citep{Quovadis.2021}.
        
        Last, the generalizability of our findings to certain other geographies is limited by the focus on one technology for hydrogen production, i.e. electrolysis. Hydrogen from steam methane reforming with carbon capture and storage represents an alternative of producing hydrogen with net neutral emissions and can be an economic alternative to electrolytic hydrogen, depending on political and geographic circumstances (see e.g. \citet{BODAL202032899}). In the German case, however, political action is strongly focused on electrolysis \citep{Wasserstoffstrategie2020}.
    
        \begin{table*}[width=2\linewidth,cols=3,pos=htbp]
            \caption{Electricity demand, wholesale price and congestion management costs in 2030}
            \label{tab:feedback_results}
            \begin{tabular}{L{5.0cm}L{3.4cm}L{3.4cm}L{3.4cm}}
                \toprule
                Scenario & Total electricity demand [TWh/year] & Mean wholesale price [EUR/MWh] & Congestion management costs [MEUR/year] \\
                \midrule
                Baseline without $H_2$& 543.90 & 62.61 & 6,163.96 \\
                \midrule
                With $H_2$, Uniform Flat Tariff & \multirow{4}{3.4cm}{616.39 (+13.33\%)} & 58.51 (-6.55\%) & 7,253.56 (+17.68\%)\\
                With $H_2$, Uniform Real-time Tariff & & 59.91 (-4.31\%) & 7,203.11 (+16.86\%)\\
                With $H_2$, Nodal Flat Tariff &  & 58.51 (-6.55\%) & 5,100.84 (-17.25\%)\\ 
                With $H_2$, Nodal Real-time Tariff &  & 59.24 (-5.38\%) & 4,915.41 (-20.26\%)\\
                \bottomrule
            \end{tabular}
        \end{table*}
        
\section{Conclusions and Policy Implications}
    Policymakers in dozens of countries are currently planning public funding for the development of future hydrogen infrastructure. They can expect that the integration of hydrogen into electricity systems will have a large effect on the operation of these systems. Our study sheds some light on the effects of hydrogen integration, and the role of spatial economic signals.\\
    
    For this, we propose a three-step methodology based on linking an electricity system dispatch model and a hydrogen supply chain model, both with granular spatial resolution. We apply this methodology for a case study of the German system in 2030.\\
    
    In the first step, we use the electricity system dispatch model to simulate uniform electricity prices -- representing current regulation -- and nodal prices, without considering hydrogen demand and production.\\ 
    
    In the second step we feed those prices into the hydrogen model, together with additional techno-economic parameters for capital and operation costs. This way, we determine the optimal spatial design of hydrogen supply chains under current uniform regulation, and a regulation with efficient spatial price signals. We identify liquefied hydrogen as the most economical form of truck based hydrogen delivery in all scenarios. Furthermore, we find that under current uniform prices, electrolyzers are cost-minimally placed close to consumption points, such as industry plants and large cities. In the nodal pricing scenario, we find that the price differences among nodes are large enough to move hydrogen production to low-cost nodes that are further away from consumption points and closer to low-cost electricity generation capacity.\\
    
    In the third step, we feed back the resulting electric loads from electrolyzers into the electricity system dispatch model. The results show that the integration of hydrogen under current uniform prices causes a large increase in congestion management costs of about 17\%, or one billion Euros per year. Thus, our analysis shows that the existing inefficiencies of single-price zonal markets can be strongly aggravated by hydrogen. Given efficient spatial economic signals, electrolyzers are integrated in a much more system-friendly way, causing a decrease in congestion management costs by up to 20\%, or about 1.1 billion Euros per year, compared to the benchmark scenario without hydrogen. When comparing the impacts of spatial (uniform vs. nodal) vs. temporal (flat vs. real-time) price resolution on congestion management costs, it becomes evident that introducing spatial variance in price signals has substantially higher benefits. The largest cost reduction can be achieved when both dimensions are combined, i.e. in a nodal real-time price signal. This is important information for policy makers in single-price electricity markets that intend to subsidize hydrogen, as our results demonstrate the considerable benefits of spatially differentiated subsidies. In fact, the subsidies a regulator would have to pay to mimic nodal prices for hydrogen within the existing single-price market design could almost entirely be covered from avoided redispatch costs.\\
    
    Given prevailing political barriers to introducing nodal pricing markets in Europe \citep{EuropeanNetworkofTransmissionSystemOperatorsforElectricity.2021} it is important to note that policy makers can still incorporate our findings within the existing single-price markets. For instance, they could design a specific nodal tariff which bills electrolyzers based on shadow nodal prices instead of wholesale prices. Alternatively, planned per-kWh subsidies (cf. \citep{Wasserstoffstrategie2020}), can be differentiated by grid node, equaling the spread between uniform prices and simulated nodal prices. Another approach could be to allow grid operators in case of congestion to curtail electrolyzers before performing the regular redispatch measures, thus creating an incentive for electrolyzer investors to avoid frequently curtailed nodes of the grid. As our study quantifies the large potential benefits of a holistic integration of hydrogen in single-price electricity systems, it also motivates future investigations into the solution space of economically efficient, and politically feasible mechanisms.\\
\printcredits

\section{Acknowledgements}
We thank Julian Huber and Marc Schmidt for their advice on efficiently deploying the electricity system dispatch model. We thank J\"urgen Beck for assistance with the procurement of hydrogen demand data.

\bibliographystyle{cas-model2-names}
\bibliography{cas-dc-template}

\begin{thebibliography}{108}
\expandafter\ifx\csname natexlab\endcsname\relax\def\natexlab#1{#1}\fi
\providecommand{\url}[1]{\texttt{#1}}
\providecommand{\href}[2]{#2}
\providecommand{\path}[1]{#1}
\providecommand{\DOIprefix}{doi:}
\providecommand{\ArXivprefix}{arXiv:}
\providecommand{\URLprefix}{URL: }
\providecommand{\Pubmedprefix}{pmid:}
\providecommand{\doi}[1]{\href{http://dx.doi.org/#1}{\path{#1}}}
\providecommand{\Pubmed}[1]{\href{pmid:#1}{\path{#1}}}
\providecommand{\bibinfo}[2]{#2}
\ifx\xfnm\relax \def\xfnm[#1]{\unskip,\space#1}\fi
\bibitem[{{Agora Energiewende} and {Wuppertal
  Institut}(2019)}]{joas2019klimaneutrale}
\bibinfo{author}{{Agora Energiewende}}, \bibinfo{author}{{Wuppertal Institut}},
  \bibinfo{year}{2019}.
\newblock \bibinfo{title}{{Klimaneutrale Industrie: Schl{\"u}sseltechnologien
  und Politikoptionen f{\"u}r Stahl, Chemie und Zement}}.
\newblock \bibinfo{journal}{Berlin, Wuppertal: Agora Energiewende} .
\bibitem[{{Antenne Brandenburg}(2020)}]{AntenneBrandenburg.2020}
\bibinfo{author}{{Antenne Brandenburg}}, \bibinfo{year}{2020}.
\newblock \bibinfo{title}{{Arcelormittal verpflichtet sich zur klimaneutralen
  Stahlproduktion}}.
\newblock \URLprefix
  \url{https://www.rbb24.de/studiofrankfurt/beitraege/2020/07/eisenhuettenstadt-arcelormittal-stahl-wasserstoff.html}.
\bibitem[{ArcelorMittal(2017)}]{ArcelorMittal.2017}
\bibinfo{author}{ArcelorMittal}, \bibinfo{year}{2017}.
\newblock \bibinfo{title}{{Wasserstoff-Stahl: ArcelorMittal und HAW Hamburg
  legen Studie vor}}.
\newblock \URLprefix
  \url{https://hamburg.arcelormittal.com/icc/arcelor-hamburg-de/broker.jsp?uMen=9d0f6fbb-a799-5199-f8b4-947d7b2f25d3&uCon=7c010c15-a102-cd51-db2a-9147d7b2f25d&uTem=aaaaaaaa-aaaa-aaaa-aaaa-000000000011}.
\bibitem[{ArcelorMittal(2020a)}]{ArcelorMittal.2020d}
\bibinfo{author}{ArcelorMittal}, \bibinfo{year}{2020}a.
\newblock \bibinfo{title}{{Hamburg: Wasserstoff-Projekt mit konkreter
  Planung}}.
\newblock \URLprefix
  \url{https://hamburg.arcelormittal.com/icc/arcelor-hamburg-de/broker.jsp?uMen=9d0f6fbb-a799-5199-f8b4-947d7b2f25d3&uCon=efe407a1-7e2f-4171-5933-6eb4ba5c485d&uTem=aaaaaaaa-aaaa-aaaa-aaaa-000000000011}.
\bibitem[{ArcelorMittal(2020b)}]{ArcelorMittal.2020c}
\bibinfo{author}{ArcelorMittal}, \bibinfo{year}{2020}b.
\newblock \bibinfo{title}{{Pressemitteilung: ArcelorMittal Europe will noch
  2020 den ersten gr{\"u}nen Stahl produzieren}}.
\newblock \URLprefix
  \url{https://bremen.arcelormittal.com/icc/arcelor-bremen-de/med/454/454205d9-81ba-1571-0202-ac33ef90c5de,11111111-1111-1111-1111-111111111111.pdf}.
\bibitem[{BASF(2020)}]{BASF.2020}
\bibinfo{author}{BASF}, \bibinfo{year}{2020}.
\newblock \bibinfo{title}{{Interview Methanpyrolyse}}.
\newblock \URLprefix
  \url{https://www.basf.com/global/de/who-we-are/sustainability/we-produce-safely-and-efficiently/energy-and-climate-protection/carbon-management/interview-methane-pyrolysis.html}.
\bibitem[{Bazzanella and Ausfelder(2017)}]{bazzanella2017low}
\bibinfo{author}{Bazzanella, A.}, \bibinfo{author}{Ausfelder, F.},
  \bibinfo{year}{2017}.
\newblock \bibinfo{title}{{Low carbon energy and feedstock for the European
  chemical industry}}.
\newblock \bibinfo{publisher}{DECHEMA, Gesellschaft f{\"u}r Chemische Technik
  und Biotechnologie eV}.
\bibitem[{Benhmad and Percebois(2018)}]{Benhmad_2018}
\bibinfo{author}{Benhmad, F.}, \bibinfo{author}{Percebois, J.},
  \bibinfo{year}{2018}.
\newblock \bibinfo{title}{Photovoltaic and wind power feed-in impact on
  electricity prices: The case of germany}.
\newblock \bibinfo{journal}{Energy Policy} \bibinfo{volume}{119},
  \bibinfo{pages}{317--326}.
\newblock \URLprefix
  \url{https://www.sciencedirect.com/science/article/pii/S0301421518302581},
  \DOIprefix\doi{https://doi.org/10.1016/j.enpol.2018.04.042}.
\bibitem[{Berger(2020)}]{Berger.2020}
\bibinfo{author}{Berger, C.}, \bibinfo{year}{2020}.
\newblock \bibinfo{title}{{95 Prozent weniger CO2 bei der Stahlproduktion}}.
\newblock \URLprefix
  \url{https://www.springerprofessional.de/anlagenbau/rohstoffe/95-prozent-weniger-co2-bei-der-stahlproduktion/18437776}.
\bibitem[{Bertsch et~al.(2015)Bertsch, Hagspiel and
  Just}]{Bertsch2015Congestion}
\bibinfo{author}{Bertsch, J.}, \bibinfo{author}{Hagspiel, S.},
  \bibinfo{author}{Just, L.}, \bibinfo{year}{2015}.
\newblock \bibinfo{title}{Congestion management in power systems: Long-term
  modeling framework and large-scale application}.
\newblock \bibinfo{type}{EWI Working Paper} \bibinfo{number}{15/03}.
  \bibinfo{address}{K\"{o}ln}.
\newblock \URLprefix \url{http://hdl.handle.net/10419/121258}.
\bibitem[{{Bezirksregierung K{\"o}ln}(2017)}]{BezirksregierungKoln.2017}
\bibinfo{author}{{Bezirksregierung K{\"o}ln}}, \bibinfo{year}{2017}.
\newblock \bibinfo{title}{{Genehmigungsbescheid: Wesentliche {\"A}nderung der
  Anlage zur Herstellung von Ammoniak (Ammoniak-Anlage) auf dem
  Werksgel{\"a}nde K{\"o}ln der Firma INEOS K{\"o}ln GmbH}}.
\newblock \URLprefix
  \url{https://www.bezreg-koeln.nrw.de/brk_internet/verfahren/52_53_industrieanlagen_genehmigungen/bekanntmachungen_koeln/ineos_koeln_gmbh_201612021/bescheid.pdf}.
\bibitem[{B{\"o}hm et~al.(2020)B{\"o}hm, Zauner, Rosenfeld and
  Tichler}]{bohm2020projecting}
\bibinfo{author}{B{\"o}hm, H.}, \bibinfo{author}{Zauner, A.},
  \bibinfo{author}{Rosenfeld, D.C.}, \bibinfo{author}{Tichler, R.},
  \bibinfo{year}{2020}.
\newblock \bibinfo{title}{Projecting cost development for future large-scale
  power-to-gas implementations by scaling effects}.
\newblock \bibinfo{journal}{Applied Energy} \bibinfo{volume}{264},
  \bibinfo{pages}{114780}.
\bibitem[{BP(2019)}]{BP.2019}
\bibinfo{author}{BP}, \bibinfo{year}{2019}.
\newblock \bibinfo{title}{{Zahlen {\&} Fakten: Produktion}}.
\newblock \URLprefix
  \url{https://www.bp.com/de_de/germany/home/wo-wir-sind/raffinerie-gelsenkirchen/wer-wir-sind/zahlen-und-fakten.html}.
\bibitem[{Brown et~al.(2018)Brown, Schlachtberger, Kies, Schramm and
  Greiner}]{Brown2018}
\bibinfo{author}{Brown, T.}, \bibinfo{author}{Schlachtberger, D.},
  \bibinfo{author}{Kies, A.}, \bibinfo{author}{Schramm, S.},
  \bibinfo{author}{Greiner, M.}, \bibinfo{year}{2018}.
\newblock \bibinfo{title}{Synergies of sector coupling and transmission
  reinforcement in a cost-optimised, highly renewable european energy system}.
\newblock \bibinfo{journal}{Energy} \bibinfo{volume}{160},
  \bibinfo{pages}{720--739}.
\newblock \DOIprefix\doi{https://doi.org/10.1016/j.energy.2018.06.222.}
\bibitem[{Br\"ummer et~al.(2021)Br\"ummer, Heim, Moser and
  Lukas}]{Quovadis.2021}
\bibinfo{author}{Br\"ummer, T.}, \bibinfo{author}{Heim, A.},
  \bibinfo{author}{Moser, H.}, \bibinfo{author}{Lukas, W.},
  \bibinfo{year}{2021}.
\newblock \bibinfo{title}{{Quo vadis, Elektrolyse?}}
\newblock \bibinfo{howpublished}{Retrieved from
  \url{https://www.element-eins.eu/_Resources/Persistent/ca8686dd02b383a73ff56cd160bdbb139dc846ed/Quo-Vadis-Elektrolyse_DIN-A4_quer_V8_download.pdf}}.
\bibitem[{Bundesnetzagentur(2018)}]{Bundesnetzagentur.2018}
\bibinfo{author}{Bundesnetzagentur}, \bibinfo{year}{2018}.
\newblock \bibinfo{title}{Marktdaten}.
\newblock \bibinfo{howpublished}{Retrieved from
  \url{https://www.smard.de/home/downloadcenter/download_marktdaten/726\#!?downloadAttributes=\%7B\%22selectedCategory\%22:1,\%22selectedSubCategory\%22:8,\%22selectedRegion\%22:\%22DE\%22,\%22from\%22:1546297200000,\%22to\%22:1577833199999,\%22selectedFileType\%22:\%22CSV\%22\%7D}}.
\bibitem[{Bundesnetzagentur(2019a)}]{NEPBest.2019}
\bibinfo{author}{Bundesnetzagentur}, \bibinfo{year}{2019}a.
\newblock \bibinfo{title}{{Best{\"a}tigung des Netzentwicklungsplans Strom
  f{\"u}r das Zieljahr 2030}}.
\newblock \bibinfo{howpublished}{Retrieved from
  \url{https://www.netzentwicklungsplan.de/sites/default/files/paragraphs-files/NEP2019-2030_Bestaetigung.pdf}}.
\bibitem[{Bundesnetzagentur(2019b)}]{BNetzA-power-plant-list}
\bibinfo{author}{Bundesnetzagentur}, \bibinfo{year}{2019}b.
\newblock \bibinfo{title}{Bnetza-power-plant-list}.
\newblock \bibinfo{howpublished}{Retrieved from
  \url{https://www.bundesnetzagentur.de/DE/Sachgebiete/ElektrizitaetundGas/Unternehmen_Institutionen/Versorgungssicherheit/Erzeugungskapazitaeten/Kraftwerksliste/kraftwerksliste-node.html}}.
\bibitem[{Bundesregierung(2019)}]{Bundesregierung.2019}
\bibinfo{author}{Bundesregierung}, \bibinfo{year}{2019}.
\newblock \bibinfo{title}{{Klimaschutzprogramm 2030 der Bundesregierung zur
  Umsetzung des Klimaschutzplans 2050}}.
\newblock \bibinfo{howpublished}{Retrieved from
  \url{https://www.bundesregierung.de/resource/blob/975226/1679914/e01d6bd855f09bf05cf7498e06d0a3ff/2019-10-09-klima-massnahmen-data.pdf?download=1}}.
\bibitem[{Bødal et~al.(2020)Bødal, Mallapragada, Botterud and
  Korpås}]{BODAL202032899}
\bibinfo{author}{Bødal, E.F.}, \bibinfo{author}{Mallapragada, D.},
  \bibinfo{author}{Botterud, A.}, \bibinfo{author}{Korpås, M.},
  \bibinfo{year}{2020}.
\newblock \bibinfo{title}{Decarbonization synergies from joint planning of
  electricity and hydrogen production: A texas case study}.
\newblock \bibinfo{journal}{International Journal of Hydrogen Energy}
  \bibinfo{volume}{45}, \bibinfo{pages}{32899--32915}.
\newblock \URLprefix
  \url{https://www.sciencedirect.com/science/article/pii/S0360319920335679},
  \DOIprefix\doi{https://doi.org/10.1016/j.ijhydene.2020.09.127}.
\bibitem[{Caglayan et~al.(2020)Caglayan, Weber, Heinrichs, Linßen, Robinius,
  Kukla and Stolten}]{CAGLAYAN20206793}
\bibinfo{author}{Caglayan, D.G.}, \bibinfo{author}{Weber, N.},
  \bibinfo{author}{Heinrichs, H.U.}, \bibinfo{author}{Linßen, J.},
  \bibinfo{author}{Robinius, M.}, \bibinfo{author}{Kukla, P.A.},
  \bibinfo{author}{Stolten, D.}, \bibinfo{year}{2020}.
\newblock \bibinfo{title}{Technical potential of salt caverns for hydrogen
  storage in europe}.
\newblock \bibinfo{journal}{International Journal of Hydrogen Energy}
  \bibinfo{volume}{45}, \bibinfo{pages}{6793--6805}.
\newblock \URLprefix
  \url{https://www.sciencedirect.com/science/article/pii/S0360319919347299},
  \DOIprefix\doi{https://doi.org/10.1016/j.ijhydene.2019.12.161}.
\bibitem[{{CDU, CSU und SPD}()}]{Koalitionsvertrag.2018}
\bibinfo{author}{{CDU, CSU und SPD}}, .
\bibitem[{{Deutsche {\"U}bertragungsnetzbetreiber}(2018)}]{UNB.2018}
\bibinfo{author}{{Deutsche {\"U}bertragungsnetzbetreiber}},
  \bibinfo{year}{2018}.
\newblock \bibinfo{title}{{EEG-Anlagenstammdaten}}.
\newblock \bibinfo{howpublished}{Retrieved from
  \url{https://www.netztransparenz.de/EEG/Anlagenstammdaten}}.
\bibitem[{Dillinger(2016)}]{Dillinger.2016}
\bibinfo{author}{Dillinger}, \bibinfo{year}{2016}.
\newblock \bibinfo{title}{{Produktion}}.
\newblock \URLprefix
  \url{https://www.dillinger.de/d/de/corporate/dillinger/produktion/}.
\bibitem[{Dillinger(2019)}]{Dillinger.2019}
\bibinfo{author}{Dillinger}, \bibinfo{year}{2019}.
\newblock \bibinfo{title}{{Nachhaltige Stahlproduktion an der Saar}}.
\newblock \URLprefix
  \url{https://www.dillinger.de/d/de/aktuelles/news/nachhaltige-stahlproduktion-an-der-saar-88575.shtml}.
\bibitem[{Egerer(2016)}]{Egerer2016Open}
\bibinfo{author}{Egerer, J.}, \bibinfo{year}{2016}.
\newblock \bibinfo{title}{Open source Electricity Model for Germany
  (ELMOD-DE)}.
\newblock \bibinfo{type}{DIW Data Documentation} \bibinfo{number}{83}. German
  Institute for Economic Research (DIW). \bibinfo{address}{Berlin}.
\newblock \URLprefix \url{http://hdl.handle.net/10419/129782}.
\bibitem[{Eicke et~al.(2020)Eicke, Khanna and Hirth}]{eicke2020locational}
\bibinfo{author}{Eicke, A.}, \bibinfo{author}{Khanna, T.},
  \bibinfo{author}{Hirth, L.}, \bibinfo{year}{2020}.
\newblock \bibinfo{title}{Locational investment signals: How to steer the
  siting of new generation capacity in power systems?}
\newblock \bibinfo{journal}{The Energy Journal} \bibinfo{volume}{41}.
\bibitem[{Emonts et~al.(2019)Emonts, Reuß, Stenzel, Welder, Knicker, Grube,
  Görner, Robinius and Stolten}]{emonts2019}
\bibinfo{author}{Emonts, B.}, \bibinfo{author}{Reuß, M.},
  \bibinfo{author}{Stenzel, P.}, \bibinfo{author}{Welder, L.},
  \bibinfo{author}{Knicker, F.}, \bibinfo{author}{Grube, T.},
  \bibinfo{author}{Görner, K.}, \bibinfo{author}{Robinius, M.},
  \bibinfo{author}{Stolten, D.}, \bibinfo{year}{2019}.
\newblock \bibinfo{title}{Flexible sector coupling with hydrogen: A
  climate-friendly fuel supply for road transport}.
\newblock \bibinfo{journal}{International Journal of Hydrogen Energy}
  \bibinfo{volume}{44}, \bibinfo{pages}{12918 -- 12930}.
\newblock \URLprefix
  \url{http://www.sciencedirect.com/science/article/pii/S0360319919312121},
  \DOIprefix\doi{https://doi.org/10.1016/j.ijhydene.2019.03.183}.
\bibitem[{{ENCON.Europe GmbH}(2018)}]{ENCON.EuropeGmbH.2018}
\bibinfo{author}{{ENCON.Europe GmbH}}, \bibinfo{year}{2018}.
\newblock \bibinfo{title}{{Potentialatlas f{\"u}r Wasserstoff: Analyse des
  Marktpotentials f{\"u}r Wasserstoff, der mit erneuerbaren Strom hergestellt
  wird, im Raffineriesektor und im zuk{\"u}nftigen Mobilit{\"a}tssektor}}.
\newblock \URLprefix
  \url{https://www.dwv-info.de/wp-content/uploads/2018/04/Potentialstudie-f\%C3\%BCr-gr\%C3\%BCnen-Wasserstoff-in-Raffinerien.pdf}.
\bibitem[{{European Alternative Fuels Observatory}()}]{VehicleDefinition}
\bibinfo{author}{{European Alternative Fuels Observatory}}, .
\newblock \bibinfo{title}{European classification for vehicle category}.
\newblock \bibinfo{howpublished}{Retrieved from
  \url{https://www.eafo.eu/knowledge-center/european-vehicle-categories}}.
\newblock \bibinfo{note}{Accessed 25.03.2020}.
\bibitem[{{European Commission}(2020)}]{EU_hydrogen_strategy}
\bibinfo{author}{{European Commission}}, \bibinfo{year}{2020}.
\newblock \bibinfo{title}{A hydrogen strategy for a climate-neutral europe}.
\newblock \bibinfo{howpublished}{Retrieved from
  \url{https://ec.europa.eu/energy/sites/ener/files/hydrogen_strategy.pdf}}.
\newblock \bibinfo{note}{Accessed 09.07.2020}.
\bibitem[{{European Network of Transmission System Operators for
  Electricity}(2018)}]{EuropeanNetworkofTransmissionSystemOperatorsforElectricity.2018}
\bibinfo{author}{{European Network of Transmission System Operators for
  Electricity}}, \bibinfo{year}{2018}.
\newblock \bibinfo{title}{{Transparency Platform}}.
\newblock \bibinfo{howpublished}{Retrieved from
  \url{https://tyndp.entsoe.eu/maps-data}}.
\bibitem[{{European Network of Transmission System Operators for
  Electricity}(2021)}]{EuropeanNetworkofTransmissionSystemOperatorsforElectricity.2021}
\bibinfo{author}{{European Network of Transmission System Operators for
  Electricity}}, \bibinfo{year}{2021}.
\newblock \bibinfo{title}{{Options for the design of European Electricity
  Markets in 2030}}.
\newblock \bibinfo{howpublished}{Retrieved from
  \url{https://eepublicdownloads.entsoe.eu/clean-documents/Publications/Market\%20Committee\%20publications/210331_Market_design\%202030.pdf}}.
\bibitem[{{FCH-JU}(2017)}]{FCH-JU2017}
\bibinfo{author}{{FCH-JU}}, \bibinfo{year}{2017}.
\newblock \bibinfo{title}{Development of business cases for fuel cells and
  hydrogen applications for regions and cities}.
\newblock \bibinfo{howpublished}{Retrieved from
  \url{https://www.fch.europa.eu/sites/default/files/171121_FCH2JU_Application-Package_WG1_Heavy\%20duty\%20trucks\%20\%28ID\%202910560\%29\%20\%28ID\%202911646\%29.pdf}}.
\newblock \bibinfo{note}{Accessed 28.03.2020}.
\bibitem[{{Federal Government of Germany}(2020)}]{Wasserstoffstrategie2020}
\bibinfo{author}{{Federal Government of Germany}}, \bibinfo{year}{2020}.
\newblock \bibinfo{title}{{Die Nationale Wasserstoffstrategie}}.
\newblock \bibinfo{howpublished}{Retrieved from
  \url{https://www.bmwi.de/Redaktion/DE/Publikationen/Energie/die-nationale-wasserstoffstrategie.pdf?__blob=publicationFile&v=12}}.
\newblock \bibinfo{note}{Accessed 10.06.2020}.
\bibitem[{Fleiter et~al.(2013)Fleiter, Schlomann and
  Eichhammer}]{fleiter2013energieverbrauch}
\bibinfo{author}{Fleiter, T.}, \bibinfo{author}{Schlomann, B.},
  \bibinfo{author}{Eichhammer, W.}, \bibinfo{year}{2013}.
\newblock \bibinfo{title}{{Energieverbrauch und CO2-Emissionen industrieller
  Prozesstechnologien: Einsparpotenziale, Hemmnisse und Instrumente}}.
\newblock \bibinfo{publisher}{Fraunhofer-Verlag}.
\newblock \URLprefix
  \url{https://www.isi.fraunhofer.de/content/dam/isi/dokumente/ccx/2013/Umweltforschungsplan_FKZ-370946130.pdf}.
\bibitem[{{Fraunhofer-Institut}(2019)}]{FraunhoferISE2019}
\bibinfo{author}{{Fraunhofer-Institut}}, \bibinfo{year}{2019}.
\newblock \bibinfo{title}{{Eine Wasserstoff-Roadmap für Deutschland}}.
\newblock \bibinfo{howpublished}{Retrieved from
  \url{https://www.ise.fraunhofer.de/content/dam/ise/de/documents/publications/studies/2019-10_Fraunhofer_Wasserstoff-Roadmap_fuer_Deutschland.pdf}}.
\newblock \bibinfo{note}{Accessed 28.03.2020}.
\bibitem[{Fraunhofer~ISI(2017)}]{BZ-LKW2017}
\bibinfo{author}{Fraunhofer~ISI, Fraunhofer~IML, P.T.C.G.},
  \bibinfo{year}{2017}.
\newblock \bibinfo{title}{{Teilstudie "Brennstoffzellen-Lkw: kritische
  Entwicklungshemmnisse, Forschungsbedarf und Marktpotential"}}.
\newblock \bibinfo{howpublished}{Retrieved from
  \url{https://www.bmvi.de/SharedDocs/DE/Anlage/G/MKS/teilstudie-brennstoffzellen-lkw.pdf?__blob=publicationFile}}.
\newblock \bibinfo{note}{Accessed 26.03.2020}.
\bibitem[{Fr{\"o}hlich et~al.(2019)Fr{\"o}hlich, Bl{\"o}mer, M{\"u}nter and
  Brischke}]{Frohlich.2019}
\bibinfo{author}{Fr{\"o}hlich, T.}, \bibinfo{author}{Bl{\"o}mer, S.},
  \bibinfo{author}{M{\"u}nter, D.}, \bibinfo{author}{Brischke, L.A.},
  \bibinfo{year}{2019}.
\newblock \bibinfo{title}{{CO2-Quellen f{\"u}r die PtX-Herstellung in
  Deutschland: Technologien, Umweltwirkung, Verf{\"u}gbarkeit}}.
\newblock \URLprefix
  \url{https://www.ifeu.de/wp-content/uploads/ifeu_paper_03_2019_CO2-Quellen-f%C3%BCr-PtX.pdf}.
\bibitem[{Golla et~al.(2020)Golla, vom Scheidt, Röhrig, Staudt and
  Weinhardt}]{Golla2020}
\bibinfo{author}{Golla, A.}, \bibinfo{author}{vom Scheidt, F.},
  \bibinfo{author}{Röhrig, N.}, \bibinfo{author}{Staudt, P.},
  \bibinfo{author}{Weinhardt, C.}, \bibinfo{year}{2020}.
\newblock \bibinfo{title}{Vehicle scheduling and refuelling of hydrogen buses
  with on-site electrolysis}.
\newblock \bibinfo{journal}{Jahrestagung der Gesellschaft f\"ur Informatik
  2020} \DOIprefix\doi{10.18420/inf2020_70}.
\bibitem[{Grimm et~al.(2019)Grimm, R{\"u}ckel, S{\"o}lch and
  Z{\"o}ttl}]{grimm2019regionally}
\bibinfo{author}{Grimm, V.}, \bibinfo{author}{R{\"u}ckel, B.},
  \bibinfo{author}{S{\"o}lch, C.}, \bibinfo{author}{Z{\"o}ttl, G.},
  \bibinfo{year}{2019}.
\newblock \bibinfo{title}{Regionally differentiated network fees to affect
  incentives for generation investment}.
\newblock \bibinfo{journal}{Energy} \bibinfo{volume}{177},
  \bibinfo{pages}{487--502}.
\bibitem[{Grube and Stolten(2018)}]{Grube2018}
\bibinfo{author}{Grube, T.}, \bibinfo{author}{Stolten, D.},
  \bibinfo{year}{2018}.
\newblock \bibinfo{title}{The impact of drive cycles and auxiliary power on
  passenger car fuel economy}.
\newblock \bibinfo{journal}{Energies} \bibinfo{volume}{11}.
\newblock \URLprefix \url{https://www.mdpi.com/1996-1073/11/4/1010},
  \DOIprefix\doi{10.3390/en11041010}.
\bibitem[{Guerra et~al.(2019)Guerra, Eichman, Kurtz and Hodge}]{GUERRA20192425}
\bibinfo{author}{Guerra, O.J.}, \bibinfo{author}{Eichman, J.},
  \bibinfo{author}{Kurtz, J.}, \bibinfo{author}{Hodge, B.M.},
  \bibinfo{year}{2019}.
\newblock \bibinfo{title}{Cost competitiveness of electrolytic hydrogen}.
\newblock \bibinfo{journal}{Joule} \bibinfo{volume}{3}, \bibinfo{pages}{2425 --
  2443}.
\newblock \URLprefix
  \url{http://www.sciencedirect.com/science/article/pii/S2542435119303228},
  \DOIprefix\doi{https://doi.org/10.1016/j.joule.2019.07.006}.
\bibitem[{{H2 MOBILITY}(2019)}]{H2Live.2019}
\bibinfo{author}{{H2 MOBILITY}}, \bibinfo{year}{2019}.
\newblock \bibinfo{title}{{Tankstellen}}.
\newblock \bibinfo{howpublished}{Retrieved from
  \url{https://h2.live/en/tankstellen}}.
\bibitem[{Hebling et~al.(2019)Hebling, Ragwitz, Fleiter, Groos, H{\"a}rle,
  Held, Jahn, M{\"u}ller, Pfeifer, Pl{\"o}tz et~al.}]{hebling2019wasserstoff}
\bibinfo{author}{Hebling, C.}, \bibinfo{author}{Ragwitz, M.},
  \bibinfo{author}{Fleiter, T.}, \bibinfo{author}{Groos, U.},
  \bibinfo{author}{H{\"a}rle, D.}, \bibinfo{author}{Held, A.},
  \bibinfo{author}{Jahn, M.}, \bibinfo{author}{M{\"u}ller, N.},
  \bibinfo{author}{Pfeifer, T.}, \bibinfo{author}{Pl{\"o}tz, P.}, et~al.,
  \bibinfo{year}{2019}.
\newblock \bibinfo{title}{{Eine Wasserstoff-Roadmap f{\"u}r Deutschland}}.
\newblock \bibinfo{journal}{Fraunhofer-Institut f{\"u}r System-und
  Innovationsforschung ISI, Karlsruhe} .
\bibitem[{Hermann et~al.(2014)Hermann, Emele and Loreck}]{oei_874}
\bibinfo{author}{Hermann, H.}, \bibinfo{author}{Emele, L.},
  \bibinfo{author}{Loreck, C.}, \bibinfo{year}{2014}.
\newblock \bibinfo{title}{{Prüfung der klimapolitischen Konsistenz und der
  Kosten von Methanisierungsstrategien}}.
\newblock \bibinfo{type}{Technical Report}. \"Oko-Institut e.V.
\newblock \URLprefix \url{https://www.oeko.de/oekodoc/2005/2014-021-de.pdf}.
\bibitem[{Heymann et~al.(2021)Heymann, Rüdisüli, vom Scheidt and
  Camanho}]{heymann.2021}
\bibinfo{author}{Heymann, F.}, \bibinfo{author}{Rüdisüli, M.},
  \bibinfo{author}{vom Scheidt, F.}, \bibinfo{author}{Camanho, A.S.},
  \bibinfo{year}{2021}.
\newblock \bibinfo{title}{Performance benchmarking of power-to-gas plants using
  composite indicators}.
\newblock \bibinfo{journal}{International Journal of Hydrogen Energy,}
  \bibinfo{volume}{forthcoming}.
\bibitem[{HKM(2020)}]{HKM.2020}
\bibinfo{author}{HKM}, \bibinfo{year}{2020}.
\newblock \bibinfo{title}{{Die Gesellschafter der HKM}}.
\newblock \URLprefix \url{https://www.hkm.de/unternehmen/gesellschafter/}.
\bibitem[{Hofbauer et~al.(2016)Hofbauer, Kaltschmitt, Keil, Neuling and
  Wagner}]{Hofbauer2016}
\bibinfo{author}{Hofbauer, H.}, \bibinfo{author}{Kaltschmitt, M.},
  \bibinfo{author}{Keil, F.}, \bibinfo{author}{Neuling, U.},
  \bibinfo{author}{Wagner, H.}, \bibinfo{year}{2016}.
\newblock \bibinfo{title}{{Vergasung in der Gasatmosph{\"a}re}}, in:
  \bibinfo{editor}{Kaltschmitt, M.}, \bibinfo{editor}{Hartmann, H.},
  \bibinfo{editor}{Hofbauer, H.} (Eds.), \bibinfo{booktitle}{{Energie aus
  Biomasse: Grundlagen, Techniken und Verfahren}}.
  \bibinfo{publisher}{Springer}, \bibinfo{address}{Berlin, Heidelberg}, pp.
  \bibinfo{pages}{1059--1182}.
\newblock \DOIprefix\doi{10.1007/978-3-662-47438-9_13}.
\bibitem[{H{\"o}lling et~al.(2017)H{\"o}lling, Weng and
  Gellert}]{holling2017bewertung}
\bibinfo{author}{H{\"o}lling, M.}, \bibinfo{author}{Weng, M.},
  \bibinfo{author}{Gellert, S.}, \bibinfo{year}{2017}.
\newblock \bibinfo{title}{{Bewertung der Herstellung von Eisenschwamm unter
  Verwendung von Wasserstoff}}.
\newblock \bibinfo{journal}{Stahl Und Eisen} \bibinfo{volume}{137},
  \bibinfo{pages}{47--53}.
\bibitem[{{Hydrogen Council} and {McKinsey \&
  Company}(2021)}]{HydrogenCouncil.2021}
\bibinfo{author}{{Hydrogen Council}}, \bibinfo{author}{{McKinsey \& Company}},
  \bibinfo{year}{2021}.
\newblock \bibinfo{title}{{Hydrogen Insights}}.
\newblock \bibinfo{howpublished}{Retrieved from
  \url{https://hydrogencouncil.com/wp-content/uploads/2021/02/Hydrogen-Insights-2021-Report.pdf}}.
\newblock \bibinfo{note}{Accessed 15.04.2021}.
\bibitem[{{Hyundai}(2020)}]{Hyundai2020}
\bibinfo{author}{{Hyundai}}, \bibinfo{year}{2020}.
\newblock \bibinfo{title}{{Erste Brennstoffzellen-Lkw Hyundai Xcient Fuel Cell
  kommen nach Europa}}.
\newblock \URLprefix
  \url{https://www.hyundai.news/de/unternehmen/erste-brennstoffzellen-lkw-hyundai-xcient-fuel-cell-kommen-nach-europa/}.
\bibitem[{{IEA}(2015)}]{H2StationStandard}
\bibinfo{author}{{IEA}}, \bibinfo{year}{2015}.
\newblock \bibinfo{title}{Large-scale hydrogen delivery infrastructure}.
\newblock \bibinfo{howpublished}{Retrieved from
  \url{http://ieahydrogen.org/Activities/Task-28/Task-28-report_final_v2_ECN_12_2_v3.aspx
  }}.
\newblock \bibinfo{note}{Accessed 28.03.2020}.
\bibitem[{IEA(2019)}]{IEA.2019}
\bibinfo{author}{IEA}, \bibinfo{year}{2019}.
\newblock \bibinfo{title}{{The Future of Hydrogen: Seizing today's
  opportunities}}.
\newblock \URLprefix
  \url{https://webstore.iea.org/download/direct/2803?fileName=The_Future_of_Hydrogen.pdf}.
\bibitem[{IKTS(2020)}]{IKTS.2020}
\bibinfo{author}{IKTS}, \bibinfo{year}{2020}.
\newblock \bibinfo{title}{{CO2-Emissionen bei der Stahlproduktion: Von 100 auf
  5 Prozent!}}
\newblock \URLprefix
  \url{https://www.ikts.fraunhofer.de/de/presse/pressemitteilungen/29-9-2020-co2-emissionen-bei-der-stahlproduktion--von-100-auf-5-.html}.
\bibitem[{IVA(2018)}]{IVA.2018}
\bibinfo{author}{IVA}, \bibinfo{year}{2018}.
\newblock \bibinfo{title}{{Wichtige Zahlen D{\"u}ngemittel: Produktion, Markt
  und Landwirtschaft: Hersteller von Ammoniak, Stickstoff- und
  Phosphat-D{\"u}ngemitteln (auf Rohphosphatbasis)}}.
\newblock \URLprefix
  \url{https://www.iva.de/sites/default/files/benutzer/%25uid/publikationen/wichtige_zahlen_2017-2018.pdf}.
\bibitem[{Jendrischik(2020)}]{Jendrischik.2020}
\bibinfo{author}{Jendrischik, M.}, \bibinfo{year}{2020}.
\newblock \bibinfo{title}{{Total setzt in Leuna auf synthetisches Methanol}}.
\newblock \URLprefix
  \url{https://www.cleanthinking.de/synthetisches-methanol-total-sunfire-wasserstoff/}.
\bibitem[{Jentsch et~al.(2014)Jentsch, Trost and Sterner}]{JENTSCH2014254}
\bibinfo{author}{Jentsch, M.}, \bibinfo{author}{Trost, T.},
  \bibinfo{author}{Sterner, M.}, \bibinfo{year}{2014}.
\newblock \bibinfo{title}{Optimal use of power-to-gas energy storage systems in
  an 85\% renewable energy scenario}.
\newblock \bibinfo{journal}{Energy Procedia} \bibinfo{volume}{46},
  \bibinfo{pages}{254--261}.
\newblock \URLprefix
  \url{https://www.sciencedirect.com/science/article/pii/S1876610214001969},
  \DOIprefix\doi{https://doi.org/10.1016/j.egypro.2014.01.180}.
  \bibinfo{note}{8th International Renewable Energy Storage Conference and
  Exhibition (IRES 2013)}.
\bibitem[{Kie{\ss}ling et~al.(2011)Kie{\ss}ling, Nefzger and
  Kaintzyk}]{kiessling2011freileitungen}
\bibinfo{author}{Kie{\ss}ling, F.}, \bibinfo{author}{Nefzger, P.},
  \bibinfo{author}{Kaintzyk, U.}, \bibinfo{year}{2011}.
\newblock \bibinfo{title}{Freileitungen: Planung, Berechnung, Ausf{\"u}hrung}.
\newblock \bibinfo{publisher}{Springer-Verlag}.
\bibitem[{Michalski et~al.(2019)Michalski, Altmann, B{\"u}nger and
  Weindorf}]{michalski2019wasserstoffstudie}
\bibinfo{author}{Michalski, M.}, \bibinfo{author}{Altmann, M.},
  \bibinfo{author}{B{\"u}nger, U.}, \bibinfo{author}{Weindorf, W.},
  \bibinfo{year}{2019}.
\newblock \bibinfo{title}{{Wasserstoffstudie Nordrhein-Westfalen}}.
\newblock \bibinfo{journal}{Eine Expertiese f{\"u}r das Ministerium f{\"u}r
  Wirtschaft, Innovation, Digitalisierung und Energie des Landes
  Nordrhein-Westfalen, D{\"u}sseldorf} .
\bibitem[{MWV(2020)}]{MWV.2020}
\bibinfo{author}{MWV}, \bibinfo{year}{2020}.
\newblock \bibinfo{title}{{Jahresbericht 2020}}.
\newblock \URLprefix
  \url{https://www.mwv.de/wp-content/uploads/2020/09/MWV_Mineraloelwirtschaftsverband-e.V.-Jahresbericht-2020-Webversion.pdf}.
\bibitem[{Neuhoff et~al.(2013)Neuhoff, Barquin, Bialek, Boyd, Dent, Echavarren,
  Grau, {von Hirschhausen}, Hobbs, Kunz, Nabe, Papaefthymiou, Weber and
  Weigt}]{NEUHOFF2013760}
\bibinfo{author}{Neuhoff, K.}, \bibinfo{author}{Barquin, J.},
  \bibinfo{author}{Bialek, J.W.}, \bibinfo{author}{Boyd, R.},
  \bibinfo{author}{Dent, C.J.}, \bibinfo{author}{Echavarren, F.},
  \bibinfo{author}{Grau, T.}, \bibinfo{author}{{von Hirschhausen}, C.},
  \bibinfo{author}{Hobbs, B.F.}, \bibinfo{author}{Kunz, F.},
  \bibinfo{author}{Nabe, C.}, \bibinfo{author}{Papaefthymiou, G.},
  \bibinfo{author}{Weber, C.}, \bibinfo{author}{Weigt, H.},
  \bibinfo{year}{2013}.
\newblock \bibinfo{title}{Renewable electric energy integration: Quantifying
  the value of design of markets for international transmission capacity}.
\newblock \bibinfo{journal}{Energy Economics} \bibinfo{volume}{40},
  \bibinfo{pages}{760--772}.
\newblock \URLprefix
  \url{https://www.sciencedirect.com/science/article/pii/S0140988313001990},
  \DOIprefix\doi{https://doi.org/10.1016/j.eneco.2013.09.004}.
\bibitem[{Nexant et~al.(2008)Nexant, Liquide, Laboratory, Venture, Institute,
  Laboratory, Laboratory and LLC}]{Nexant.2008}
\bibinfo{author}{Nexant, I.}, \bibinfo{author}{Liquide, A.},
  \bibinfo{author}{Laboratory, A.N.}, \bibinfo{author}{Venture, C.T.},
  \bibinfo{author}{Institute, G.T.}, \bibinfo{author}{Laboratory, N.R.E.},
  \bibinfo{author}{Laboratory, P.N.N.}, \bibinfo{author}{LLC, T.},
  \bibinfo{year}{2008}.
\newblock \bibinfo{title}{{Wasserstoff: Bundesregierung schlägt Kriterien zur
  Umlagebefreiung vor}}.
\newblock \bibinfo{howpublished}{Retrieved from
  \url{https://www.energate-messenger.de/news/210637/wasserstoff-bundesregierung-schlaegt-kriterien-zur-umlagebefreiung-vor}}.
\newblock \bibinfo{note}{Accessed 15.04.2021}.
\bibitem[{{OpenStreetMap Contributors}(2020)}]{OpenStreetMap}
\bibinfo{author}{{OpenStreetMap Contributors}}, \bibinfo{year}{2020}.
\newblock \bibinfo{title}{Openstreetmap data for germany}.
\newblock \bibinfo{howpublished}{Retrieved from
  \url{https://download.geofabrik.de/europe/germany.html}}.
\newblock \bibinfo{note}{Accessed 28.03.2020}.
\bibitem[{Peters and Thumann(2016)}]{Peters.2016}
\bibinfo{author}{Peters, K.}, \bibinfo{author}{Thumann, R.},
  \bibinfo{year}{2016}.
\newblock \bibinfo{title}{{Yara: Aus Dithmarschen-Wiki}}.
\newblock \URLprefix \url{https://www.dithmarschen-wiki.de/Yara}.
\bibitem[{Pototschnig(2021)}]{Pototschnig.2021}
\bibinfo{author}{Pototschnig, A.}, \bibinfo{year}{2021}.
\newblock \bibinfo{title}{{Renewable hydrogen and the 'additionality'
  requirement: why making it more complex than is needed?}}
\newblock \bibinfo{howpublished}{Retrieved from
  \url{https://cadmus.eui.eu/bitstream/handle/1814/72459/PB_2021_36_FSR.pdf?sequence=1}}.
\bibitem[{{Prognos AG}(2020a)}]{PrognosAG.2020b}
\bibinfo{author}{{Prognos AG}}, \bibinfo{year}{2020}a.
\newblock \bibinfo{title}{{Energiewirtschaftliche Projektionen und
  Folgeabsch{\"a}tzungen 2030/2050}}.
\newblock \URLprefix
  \url{https://www.bmwi.de/Redaktion/DE/Publikationen/Wirtschaft/klimagutachten.pdf?__blob=publicationFile&v=8}.
\bibitem[{{Prognos AG}(2020b)}]{PrognosAG.2020}
\bibinfo{author}{{Prognos AG}}, \bibinfo{year}{2020}b.
\newblock \bibinfo{title}{{Kosten und Transformationspfade f{\"u}r
  strombasierte Energietr{\"a}ger: Studie im Auftrag des Bundesministeriums
  f{\"u}r Wirtschaft und Energie}}.
\newblock \URLprefix
  \url{https://www.bmwi.de/Redaktion/DE/Downloads/Studien/transformationspfade-fuer-strombasierte-energietraeger.pdf?__blob=publicationFile}.
\bibitem[{Rechenberger(2020)}]{Rechenberger.2020}
\bibinfo{author}{Rechenberger, D.}, \bibinfo{year}{2020}.
\newblock \bibinfo{title}{{BASF SE Produktion Ammoniak: E-Mail}}.
\bibitem[{Redenius(2020a)}]{Redenius.2020}
\bibinfo{author}{Redenius, A.}, \bibinfo{year}{2020}a.
\newblock \bibinfo{title}{{SALCOS, WindH2, GrInHy -- Wasserstoffprojekte bei
  der Salzgitter AG}}.
\newblock \URLprefix
  \url{https://www.prozesswaerme.net/fileadmin/Prozesswaerme/Dateien_Redaktion/Ausgewaehlte_Beitraege/13_sonderteil_salzgitter.pdf}.
\bibitem[{Redenius(2020b)}]{Redenius.2020b}
\bibinfo{author}{Redenius, A.}, \bibinfo{year}{2020}b.
\newblock \bibinfo{title}{{Vortrag auf der WerkstoffPlusAuto 2020: SALCOS -
  nachhaltige, flexible, CO2-arme Stahlproduktion}}.
\bibitem[{Reuß(2019)}]{Reuss2019Dissertation}
\bibinfo{author}{Reuß, M.}, \bibinfo{year}{2019}.
\newblock \bibinfo{title}{Techno-ökonomische Analyse alternativer
  Wasserstoffinfrastruktur}.
\newblock Ph.D. thesis. RWTH Aachen.
\newblock \DOIprefix\doi{10.18154/RWTH-2019-07432}.
\bibitem[{Reuß et~al.(2019)Reuß, Grube, Robinius and Stolten}]{Reuss2019}
\bibinfo{author}{Reuß, M.}, \bibinfo{author}{Grube, T.},
  \bibinfo{author}{Robinius, M.}, \bibinfo{author}{Stolten, D.},
  \bibinfo{year}{2019}.
\newblock \bibinfo{title}{A hydrogen supply chain with spatial resolution:
  Comparative analysis of infrastructure technologies in germany}.
\newblock \bibinfo{journal}{Applied Energy} \bibinfo{volume}{247},
  \bibinfo{pages}{438--453}.
\newblock \DOIprefix\doi{https://doi.org/10.1016/j.apenergy.2019.04.064}.
\bibitem[{Robinius et~al.(2017)Robinius, Otto, Syranidis, Ryberg, Heuser,
  Welder, Grube, Markewitz, Tietze and Stolten}]{Robinius_2017_Part2}
\bibinfo{author}{Robinius, M.}, \bibinfo{author}{Otto, A.},
  \bibinfo{author}{Syranidis, K.}, \bibinfo{author}{Ryberg, D.S.},
  \bibinfo{author}{Heuser, P.}, \bibinfo{author}{Welder, L.},
  \bibinfo{author}{Grube, T.}, \bibinfo{author}{Markewitz, P.},
  \bibinfo{author}{Tietze, V.}, \bibinfo{author}{Stolten, D.},
  \bibinfo{year}{2017}.
\newblock \bibinfo{title}{Linking the power and transport sectors—part 2:
  Modelling a sector coupling scenario for germany}.
\newblock \bibinfo{journal}{Energies} \bibinfo{volume}{10}.
\newblock \URLprefix \url{https://www.mdpi.com/1996-1073/10/7/957},
  \DOIprefix\doi{10.3390/en10070957}.
\bibitem[{ROGESA(2016)}]{ROGESA.2016}
\bibinfo{author}{ROGESA}, \bibinfo{year}{2016}.
\newblock \bibinfo{title}{{Daten und Fakten}}.
\newblock \URLprefix
  \url{http://www.rogesa.de/rogesa/produktion/daten/index.shtml.de}.
\bibitem[{Rose and Neumann(2020a)}]{Rose2020}
\bibinfo{author}{Rose, P.K.}, \bibinfo{author}{Neumann, F.},
  \bibinfo{year}{2020}a.
\newblock \bibinfo{title}{Hydrogen refueling station networks for heavy-duty
  vehicles in future power systems}.
\newblock \bibinfo{journal}{Transportation Research Part D: Transport and
  Environment} \bibinfo{volume}{83}, \bibinfo{pages}{102358}.
\newblock \URLprefix
  \url{http://www.sciencedirect.com/science/article/pii/S1361920920305459},
  \DOIprefix\doi{https://doi.org/10.1016/j.trd.2020.102358}.
\bibitem[{Rose and Neumann(2020b)}]{Rose&Neumann2020}
\bibinfo{author}{Rose, P.K.}, \bibinfo{author}{Neumann, F.},
  \bibinfo{year}{2020}b.
\newblock \bibinfo{title}{Hydrogen refueling station networks for heavy-duty
  vehicles in future power systems}.
\newblock \bibinfo{journal}{Transportation Research Part D: Transport and
  Environment} \bibinfo{volume}{83}.
\newblock \DOIprefix\doi{https://doi.org/10.1016/j.trd.2020.102358}.
\bibitem[{Ruhnau(2020)}]{Ruhnau2020Market}
\bibinfo{author}{Ruhnau, O.}, \bibinfo{year}{2020}.
\newblock \bibinfo{title}{Market-based renewables: How flexible hydrogen
  electrolyzers stabilize wind and solar market values}.
\newblock \bibinfo{type}{Technical Report}. Hertie School.
  \bibinfo{address}{Kiel, Hamburg}.
\newblock \URLprefix \url{http://hdl.handle.net/10419/227075}.
\bibitem[{Runge et~al.(2019)Runge, Sölch, Albert, Wasserscheid, Zöttl and
  Grimm}]{Runge2019}
\bibinfo{author}{Runge, P.}, \bibinfo{author}{Sölch, C.},
  \bibinfo{author}{Albert, J.}, \bibinfo{author}{Wasserscheid, P.},
  \bibinfo{author}{Zöttl, G.}, \bibinfo{author}{Grimm, V.},
  \bibinfo{year}{2019}.
\newblock \bibinfo{title}{Economic comparison of different electric fuels for
  energy scenarios in 2035}.
\newblock \bibinfo{journal}{Applied Energy} \bibinfo{volume}{233–234},
  \bibinfo{pages}{1078--1093}.
\newblock \DOIprefix\doi{https://doi.org/10.1016/j.apenergy.2018.10.023.}
\bibitem[{Runge et~al.(2020)Runge, Sölch, Albert, Wasserscheid, Zöttl and
  Grimm}]{runge2020}
\bibinfo{author}{Runge, P.}, \bibinfo{author}{Sölch, C.},
  \bibinfo{author}{Albert, J.}, \bibinfo{author}{Wasserscheid, P.},
  \bibinfo{author}{Zöttl, G.}, \bibinfo{author}{Grimm, V.},
  \bibinfo{year}{2020}.
\newblock \bibinfo{title}{Economic comparison of electric fuels produced at
  excellent locations for renewable energies: A scenario for 2035}.
\newblock \bibinfo{journal}{Working paper}
  \DOIprefix\doi{http://dx.doi.org/10.2139/ssrn.3623514}.
\bibitem[{vom Scheidt et~al.(2020)vom Scheidt, Medinov{\'a}, Ludwig, Richter,
  Staudt and Weinhardt}]{vomScheidt_2020_Review}
\bibinfo{author}{vom Scheidt, F.}, \bibinfo{author}{Medinov{\'a}, H.},
  \bibinfo{author}{Ludwig, N.}, \bibinfo{author}{Richter, B.},
  \bibinfo{author}{Staudt, P.}, \bibinfo{author}{Weinhardt, C.},
  \bibinfo{year}{2020}.
\newblock \bibinfo{title}{Data analytics in the electricity sector--a
  quantitative and qualitative literature review}.
\newblock \bibinfo{journal}{Energy and AI} , \bibinfo{pages}{100009}.
\bibitem[{Schmidt and Zinke(2020)}]{Schmidt2020price}
\bibinfo{author}{Schmidt, L.}, \bibinfo{author}{Zinke, J.},
  \bibinfo{year}{2020}.
\newblock \bibinfo{title}{One price fits all? Wind power expansion under
  uniform and nodal pricing in Germany}.
\newblock \bibinfo{type}{EWI Working Paper} \bibinfo{number}{20/06}.
  \bibinfo{address}{Cologne}.
\newblock \URLprefix \url{http://hdl.handle.net/10419/227510}.
\bibitem[{Schmidt et~al.(2017)Schmidt, Gambhir, Staffell, Hawkes, Nelson and
  Few}]{Schmidt2017}
\bibinfo{author}{Schmidt, O.}, \bibinfo{author}{Gambhir, A.},
  \bibinfo{author}{Staffell, I.}, \bibinfo{author}{Hawkes, A.},
  \bibinfo{author}{Nelson, J.}, \bibinfo{author}{Few, S.},
  \bibinfo{year}{2017}.
\newblock \bibinfo{title}{Future cost and performance of water electrolysis: An
  expert elicitation study}.
\newblock \bibinfo{journal}{International Journal of Hydrogen Energy}
  \bibinfo{volume}{42}, \bibinfo{pages}{30470--30492}.
\newblock \DOIprefix\doi{https://doi.org/10.1016/j.ijhydene.2017.10.045.}
\bibitem[{Schweer et~al.(2002)Schweer, Scholz and Heisel}]{schweer2002site}
\bibinfo{author}{Schweer, D.}, \bibinfo{author}{Scholz, G.},
  \bibinfo{author}{Heisel, M.}, \bibinfo{year}{2002}.
\newblock \bibinfo{title}{{On-site-Versorgung von Erd{\"o}lraffinerien mit
  technischen Gasen}}.
\newblock \bibinfo{journal}{Erd{\"o}l, Erdgas, Kohle} \bibinfo{volume}{118},
  \bibinfo{pages}{115--120}.
\bibitem[{{skw Piesteritz}(2015)}]{skwPiesteritz.2015}
\bibinfo{author}{{skw Piesteritz}}, \bibinfo{year}{2015}.
\newblock \bibinfo{title}{{Sicherheitsdatenblatt: Ammoniak}}.
\newblock \URLprefix \url{https://www.skwp.de/media-center/broschueren/reach/}.
\bibitem[{Stagge(2020)}]{Stagge.2020}
\bibinfo{author}{Stagge, M.}, \bibinfo{year}{2020}.
\newblock \bibinfo{title}{{Wasserstoffbedarf thyssenkrupp 2030: E-Mail}}.
\bibitem[{Staudt(2019)}]{staudt2019transmission}
\bibinfo{author}{Staudt, P.}, \bibinfo{year}{2019}.
\newblock \bibinfo{title}{Transmission Congestion Management in Electricity
  Grids-Designing Markets and Mechanisms}.
\newblock Ph.D. thesis. KIT-Bibliothek.
\bibitem[{{Staudt} and {Oren}(2020)}]{Staudt2020}
\bibinfo{author}{{Staudt}, P.}, \bibinfo{author}{{Oren}, S.},
  \bibinfo{year}{2020}.
\newblock \bibinfo{title}{A merchant transmission approach for uniform-price
  electricity markets}, in: \bibinfo{booktitle}{Proceedings of the 53rd Hawaii
  International Conference on System Sciences}, pp. \bibinfo{pages}{1--10}.
\bibitem[{Staudt et~al.(2019)Staudt, Rausch, Gärttner and
  Weinhardt}]{Staudt_2019}
\bibinfo{author}{Staudt, P.}, \bibinfo{author}{Rausch, B.},
  \bibinfo{author}{Gärttner, J.}, \bibinfo{author}{Weinhardt, C.},
  \bibinfo{year}{2019}.
\newblock \bibinfo{title}{Predicting transmission line congestion in energy
  systems with a high share of renewables}, in: \bibinfo{booktitle}{2019 IEEE
  Milan PowerTech}, pp. \bibinfo{pages}{1--6}.
\newblock \DOIprefix\doi{10.1109/PTC.2019.8810527}.
\bibitem[{{Staudt} et~al.(2017){Staudt}, {Wegner}, {Garttner} and
  {Weinhardt}}]{Staudt2017}
\bibinfo{author}{{Staudt}, P.}, \bibinfo{author}{{Wegner}, F.},
  \bibinfo{author}{{Garttner}, J.}, \bibinfo{author}{{Weinhardt}, C.},
  \bibinfo{year}{2017}.
\newblock \bibinfo{title}{Analysis of redispatch and transmission capacity
  pricing on a local electricity market setup}, in: \bibinfo{booktitle}{2017
  14th International Conference on the European Energy Market (EEM)}, pp.
  \bibinfo{pages}{1--6}.
\bibitem[{swb(2020)}]{swb.2020}
\bibinfo{author}{swb}, \bibinfo{year}{2020}.
\newblock \bibinfo{title}{{Nachhaltigkeit: Wasserstoff: Mit gr{\"u}nem Stahl
  Emissionen reduzieren}}.
\newblock \URLprefix
  \url{https://www.swb.de/ueber-swb/unternehmen/nachhaltigkeit/wasserstoff/elektrolyseur}.
\bibitem[{Thoma(2020)}]{Thoma.2020}
\bibinfo{author}{Thoma, H.J.}, \bibinfo{year}{2020}.
\newblock \bibinfo{title}{{Methanol Produktionskapazit{\"a}t: E-Mail}}.
\bibitem[{thyssenkrupp(2019)}]{thyssenkrupp.2019}
\bibinfo{author}{thyssenkrupp}, \bibinfo{year}{2019}.
\newblock \bibinfo{title}{{Wasserstoff statt Kohle. thyssenkrupp Steel startet
  wegweisendes Projekt f{\"u}r eine klimafreundliche Stahlproduktion am
  Standort Duisburg}}.
\newblock \URLprefix
  \url{https://www.thyssenkrupp-steel.com/de/newsroom/pressemitteilungen/wasserstoff-statt-kohle.html}.
\bibitem[{thyssenkrupp(2020a)}]{thyssenkrupp.2020}
\bibinfo{author}{thyssenkrupp}, \bibinfo{year}{2020}a.
\newblock \bibinfo{title}{{Die Klimastrategie von thyssenkrupp Steel zur
  nachhaltigen Stahlproduktion}}.
\newblock \URLprefix
  \url{https://www.thyssenkrupp-steel.com/de/unternehmen/nachhaltigkeit/klimastrategie/}.
\bibitem[{thyssenkrupp(2020b)}]{thyssenkrupp.2020b}
\bibinfo{author}{thyssenkrupp}, \bibinfo{year}{2020}b.
\newblock \bibinfo{title}{{Gr{\"u}ner Wasserstoff f{\"u}r die Stahlproduktion:
  RWE und thyssenkrupp planen Zusammenarbeit}}.
\newblock \URLprefix
  \url{https://www.thyssenkrupp-steel.com/media/content_1/presse/dokumente/2020_1/juni_2/20200610_pm_thyssenkrupp_steel_loi_rwe_pre-final.pdf}.
\bibitem[{Tlili et~al.(2020)Tlili, Mansilla, Linßen, Reuß, Grube, Robinius,
  André, Perez, {Le Duigou} and Stolten}]{TLILI.2020}
\bibinfo{author}{Tlili, O.}, \bibinfo{author}{Mansilla, C.},
  \bibinfo{author}{Linßen, J.}, \bibinfo{author}{Reuß, M.},
  \bibinfo{author}{Grube, T.}, \bibinfo{author}{Robinius, M.},
  \bibinfo{author}{André, J.}, \bibinfo{author}{Perez, Y.},
  \bibinfo{author}{{Le Duigou}, A.}, \bibinfo{author}{Stolten, D.},
  \bibinfo{year}{2020}.
\newblock \bibinfo{title}{Geospatial modelling of the hydrogen infrastructure
  in france in order to identify the most suited supply chains}.
\newblock \bibinfo{journal}{International Journal of Hydrogen Energy}
  \bibinfo{volume}{45}, \bibinfo{pages}{3053--3072}.
\newblock \URLprefix
  \url{https://www.sciencedirect.com/science/article/pii/S0360319919341503},
  \DOIprefix\doi{https://doi.org/10.1016/j.ijhydene.2019.11.006}.
\bibitem[{VCI(2020)}]{VCI.2020}
\bibinfo{author}{VCI}, \bibinfo{year}{2020}.
\newblock \bibinfo{title}{{Chemiewirtschaft in Zahlen 2020}}.
\newblock \URLprefix
  \url{https://www.vci.de/vci/downloads-vci/publikation/chemiewirtschaft-in-zahlen-print.pdf}.
\bibitem[{Victoria et~al.(2019)Victoria, Zhu, Brown, Andresen and
  Greiner}]{VICTORIA2019111977}
\bibinfo{author}{Victoria, M.}, \bibinfo{author}{Zhu, K.},
  \bibinfo{author}{Brown, T.}, \bibinfo{author}{Andresen, G.B.},
  \bibinfo{author}{Greiner, M.}, \bibinfo{year}{2019}.
\newblock \bibinfo{title}{The role of storage technologies throughout the
  decarbonisation of the sector-coupled european energy system}.
\newblock \bibinfo{journal}{Energy Conversion and Management}
  \bibinfo{volume}{201}, \bibinfo{pages}{111977}.
\newblock \URLprefix
  \url{https://www.sciencedirect.com/science/article/pii/S0196890419309835},
  \DOIprefix\doi{https://doi.org/10.1016/j.enconman.2019.111977}.
\bibitem[{{vom Scheidt} et~al.(2020){vom Scheidt}, M\"uller, Staudt and
  Weinhardt}]{data2030}
\bibinfo{author}{{vom Scheidt}, F.}, \bibinfo{author}{M\"uller, C.},
  \bibinfo{author}{Staudt, P.}, \bibinfo{author}{Weinhardt, C.},
  \bibinfo{year}{2020}.
\newblock \bibinfo{title}{{The German electricity system in 2030: data on
  consumption, generation, and the grid}}.
\newblock \DOIprefix\doi{10.5445/IR/1000125576}.
\bibitem[{{vom Scheidt} et~al.(2021){vom Scheidt}, Qu, Staudt, Mallapragada and
  Weinhardt}]{vomscheidt.2021}
\bibinfo{author}{{vom Scheidt}, F.}, \bibinfo{author}{Qu, J.},
  \bibinfo{author}{Staudt, P.}, \bibinfo{author}{Mallapragada, D.},
  \bibinfo{author}{Weinhardt, C.}, \bibinfo{year}{2021}.
\newblock \bibinfo{title}{{The effects of electricity tariffs on cost-minimal
  hydrogen supply chains and their impact on electricity prices and redispatch
  costs}}.
\newblock \bibinfo{journal}{{Proceedings of the 54th Hawaii International
  Conference on System Sciences}} \URLprefix
  \url{http://hdl.handle.net/10125/71016}.
\bibitem[{Warscheid(2020a)}]{Warscheid.b}
\bibinfo{author}{Warscheid, L.}, \bibinfo{year}{2020}a.
\newblock \bibinfo{title}{{Fahrplan der Saar-Stahlkocher zum gr{\"u}nen
  Stahl}}.
\newblock \bibinfo{journal}{Saarbruecker Zeitung} \bibinfo{volume}{2020}.
\newblock \URLprefix
  \url{https://www.saarbruecker-zeitung.de/saarland/saar-wirtschaft/stufenplan-zu-gruenem-stahl-aus-dem-saarland-bis-2050_aid-53012835}.
\bibitem[{Warscheid(2020b)}]{Warscheid.}
\bibinfo{author}{Warscheid, L.}, \bibinfo{year}{2020}b.
\newblock \bibinfo{title}{{In Dillingen wird jetzt mit Wasserstoff gekocht}}.
\newblock \bibinfo{journal}{Saarbruecker Zeitung} \bibinfo{volume}{2020}.
\newblock \URLprefix
  \url{https://www.saarbruecker-zeitung.de/saarland/saar-wirtschaft/neue-anlage-von-rogesa-saarstahl-dillinger-nutzt-wasserstoff_aid-52906817?utm_source=mail&utm_medium=referral&utm_campaign=share#successLogin}.
\bibitem[{Weger et~al.(2020)Weger, Leitão and Lawrence}]{WEGER2020}
\bibinfo{author}{Weger, L.B.}, \bibinfo{author}{Leitão, J.},
  \bibinfo{author}{Lawrence, M.G.}, \bibinfo{year}{2020}.
\newblock \bibinfo{title}{Expected impacts on greenhouse gas and air pollutant
  emissions due to a possible transition towards a hydrogen economy in german
  road transport}.
\newblock \bibinfo{journal}{International Journal of Hydrogen Energy}
  \URLprefix
  \url{http://www.sciencedirect.com/science/article/pii/S0360319920341847},
  \DOIprefix\doi{https://doi.org/10.1016/j.ijhydene.2020.11.014}.
\bibitem[{Wilms et~al.(2018)Wilms, Sch{\"a}fer-Stradowsky and
  Jahnke}]{wilms2018heutige}
\bibinfo{author}{Wilms, S.}, \bibinfo{author}{Sch{\"a}fer-Stradowsky, S.},
  \bibinfo{author}{Jahnke, P.}, \bibinfo{year}{2018}.
\newblock \bibinfo{title}{{Heutige Einsatzgebiete f{\"u}r Power
  Fuels—Factsheets zur Anwendung von klimafreundlich erzeugten synthetischen
  Energietr{\"a}gern}}.
\newblock \URLprefix
  \url{https://www.dena.de/fileadmin/dena/Publikationen/PDFs/2019/181123_dena_PtX-Factsheets.pdf}.
\bibitem[{{WV Stahl}(2020)}]{stahl2020statistisches}
\bibinfo{author}{{WV Stahl}}, \bibinfo{year}{2020}.
\newblock \bibinfo{title}{{Statistisches Jahrbuch der Stahlindustrie
  2019/2020}}.
\bibitem[{Xiong et~al.(2021)Xiong, Predel, {Crespo del Granado} and
  Egging-Bratseth}]{XIONG2021116201}
\bibinfo{author}{Xiong, B.}, \bibinfo{author}{Predel, J.},
  \bibinfo{author}{{Crespo del Granado}, P.}, \bibinfo{author}{Egging-Bratseth,
  R.}, \bibinfo{year}{2021}.
\newblock \bibinfo{title}{Spatial flexibility in redispatch: Supporting low
  carbon energy systems with power-to-gas}.
\newblock \bibinfo{journal}{Applied Energy} \bibinfo{volume}{283},
  \bibinfo{pages}{116201}.
\newblock \URLprefix
  \url{https://www.sciencedirect.com/science/article/pii/S0306261920315981},
  \DOIprefix\doi{https://doi.org/10.1016/j.apenergy.2020.116201}.
\bibitem[{Yilmaz(2018)}]{yilmaz2018massnahmen}
\bibinfo{author}{Yilmaz, C.}, \bibinfo{year}{2018}.
\newblock \bibinfo{title}{{Ma{\ss}nahmen zur Dekarbonisierung des
  Hochofenprozesses durch Einsatz von Wasserstoff}}.
\newblock \bibinfo{publisher}{Cuvillier Verlag}.
\bibitem[{Zhang et~al.(2020)Zhang, Greenblatt, Wei, Eichman, Saxena, Muratori
  and Guerra}]{ZHANG2020115651}
\bibinfo{author}{Zhang, C.}, \bibinfo{author}{Greenblatt, J.B.},
  \bibinfo{author}{Wei, M.}, \bibinfo{author}{Eichman, J.},
  \bibinfo{author}{Saxena, S.}, \bibinfo{author}{Muratori, M.},
  \bibinfo{author}{Guerra, O.J.}, \bibinfo{year}{2020}.
\newblock \bibinfo{title}{Flexible grid-based electrolysis hydrogen production
  for fuel cell vehicles reduces costs and greenhouse gas emissions}.
\newblock \bibinfo{journal}{Applied Energy} \bibinfo{volume}{278},
  \bibinfo{pages}{115651}.
\newblock \URLprefix
  \url{https://www.sciencedirect.com/science/article/pii/S0306261920311491},
  \DOIprefix\doi{https://doi.org/10.1016/j.apenergy.2020.115651}.

\end{thebibliography}

\appendix
\section{Appendix A: Hydrogen demand data} \label{appendix_A}

    \paragraph{\textbf{Steel}}
        To identify all steel plants with potential for hydrogen use in 2030, we use the statistical report of the steel industry \citep{stahl2020statistisches}. Looking at future hydrogen demand, only those 70 \% of steel producers who manufacture via the blast furnace route are relevant, as large quantities of CO\textsubscript{2} are emitted here and can be avoided by switching to the direct reduction route. In addition, the ArcelorMittal plant in Hamburg is included, as it already uses a direct reduction approach \citep{holling2017bewertung}. Table \ref{tab:steel_demand} lists the eight identified steel production sites.
        
        The production volumes and relative shares of primary and secondary steel in Germany have been approximately constant since 2012 \citep{stahl2020statistisches}. Therefore, and in line with \cite{hebling2019wasserstoff}, we use past production volume and distribution as 2030 estimates. In particular, we use 2017 values, as only those are available in \citep{stahl2020statistisches}. Table \ref{tab:steel_demand} shows the crude steel quantities produced in 2017 for each identified site with potential hydrogen demand.
        
        However, it can be assumed that not all steel producers will switch to direct reduction by 2030, due to various reasons. For instance, the switch is associated with high investment costs, is technically demanding \citep{IKTS.2020}, and comes with new uncertainties like future hydrogen costs \citep{joas2019klimaneutrale}. Correspondingly, steel producers are planning individual solutions for medium-term CO\textsubscript{2} emission reduction to achieve reduction goals. Therefore, all relevant plants must be analyzed individually.
    
        ArcelorMittal Hamburg has been operating a direct reduction plant since the mid-1970s \citep{holling2017bewertung}. The reduction gas used today consists of about 60 \% hydrogen \citep{ArcelorMittal.2017}. By 2030, steel production is planned to be completely CO\textsubscript{2}-neutral \citep{ArcelorMittal.2020d}. Accordingly, we assume that there will be a complete switch to the direct reduction route with 100 \% hydrogen input by 2030. For the direct reduction route, we assume the specific hydrogen demand factor $80 kg_{H_2}/t_{steel}$, based on \cite{michalski2019wasserstoffstudie}. The hydrogen demand of ArcelorMittal Hamburg for the year 2030 is estimated with equation \ref{equ:steel}.
        
        \begin{equation}
        \label{equ:steel}
                \begin{split} 
                    HD = Output_{t_{Steel}} * specificDemandFactor \\ * 33.33 kWh_{H_2}/kg_{H_2} 
                \end{split} 
        \end{equation}
        
        ArcelorMittal Eisenhüttenstadt and ArcelorMittal Duisburg have not publicly announced any plans to use hydrogen until 2030, but it has been indicated that long-term adoption of hydrogen for the former plant will depend on the results of current pilot projects of the ArcelorMittal group \citep{AntenneBrandenburg.2020, ArcelorMittal.2020c}. Therefore, we assume that these plants do not have any hydrogen demand in 2030.
        
        ArcelorMittal Bremen is focusing on the use of hydrogen via the blast furnace route to achieve the medium-term goals. However, the company already plans to construct an electrolyser on-site \citep{swb.2020} that will be sufficient to fully meet the hydrogen demand in 2030. Thus, the plant does not have any net demand for hydrogen.
        
        ROGESA, a subsidiary of Dillinger and Saarstahl, produces pig iron, which is supplied to Dillinger and Saarstahl for the subsequent crude steel production \citep{Dillinger.2016}. Therefore, Dillinger and Saarstahl are considered collectively for further calculations. ROGESA operates two blast furnaces and plans to optimise both by blowing in hydrogen as a reducing agent in order to achieve a reduction in CO\textsubscript{2} emissions \citep{Dillinger.2019}. According to a step-by-step plan of the Saarland-based steel industry, both blast furnaces are to remain in operation until 2031 \citep{Warscheid.b}. In addition, an electric furnace and a direct reduction plant are to be built, which will initially only use natural gas to produce directly reduced iron from iron ore \citep{Warscheid.b}. Therefore, we assume that by 2030, both blast furnaces will use the maximum amount of hydrogen. Both blast furnaces are technically able to use a maximum of approximately 3,700 $kg_{H_2}/h$ \citep{Warscheid., ROGESA.2016}. Thus, the hydrogen demand of ROGESA (Dillinger and Saarstahl) for the year 2030 is estimated to be 2.1606 $TWh_{H_2}$, based on equation \ref{equ:rogesa}. \footnote{To validate the results, we also estimate the demand with equation \ref{equ:steel}, which returns 2.0668 $TWh_{H_2}$ and thus confirms the calculations. For all further calculations, we use 2.1606 $TWh_{H_2}$ as demand for the Dillinger and Saarstahl steel plants.}
        \begin{equation}
        \label{equ:rogesa}
        \begin{split}
            HD = 2 * 3700kg_{H_2}/h * 8760h \\ * 33.33 kWh_{H_2}/kg_{H_2} 
        \end{split}
        \end{equation}
        
        Hüttenwerke Krupp Mannesmann (HKM) is owned 50 \% by Thyssenkrupp Steel Europe AG, 30 \% by Salzgitter Mannesmann GmbH and 20 \% by the French company Vallourec Tubes S.A.S \citep{HKM.2020}. Regarding the use of hydrogen in production, no press reports were found that were published by HKM. Consequently, it is assumed that due to the structure of the company, no hydrogen will be used until 2030, as the shareholders might primarily concentrate on their own production facilities and their optimisation.
        
        Salzgitter is pursuing a gradual conversion to hydrogen-based steel production via the direct reduction/electric arc furnace route. In the first stage of expansion, a direct reduction plant and an electric arc furnace will be built \citep{Redenius.2020}. This expansion stage will lead to a hydrogen use of 81,332 Nm\textsuperscript{3}/h and a specific hydrogen demand factor of 12.27 $kg_{H_2}/t_{steel}$ \citep{Redenius.2020b} for the overall plant output. Thus, the hydrogen demand for 2030 can be calculated with equation \ref{equ:steel}.
        
        Thyssenkrupp plans to replace two blast furnaces with two direct reduction plants, and to optimize one blast furnace by blowing in hydrogen until 2030 \citep{thyssenkrupp.2020}. Current estimations indicate that around 200,000 tons of hydrogen per year will be needed from 2030 \citep{Stagge.2020}. A share of this will be supplied through a long-term contract with RWE, from a 100 MW electrolyzer capable of supplying 1.7 tons of hydrogen per hour \citep{thyssenkrupp.2020b}. This supply is deducted from the total demand in order to calculate the hydrogen net demand for 2030 as shown in equation \ref{equ:thyssenkrupp}.
        \begin{equation}
        \label{equ:thyssenkrupp}
        \begin{split}
            HD = (200,000,000kg_{H_2} - 1,700kg_{H_2}/h \\ * 8,760h) * 33.33 kWh_{H_2}/kg_{H_2}  
        \end{split}
        \end{equation}
            
    \paragraph{\textbf{Ammonia}}
        The hydrogen demand from the German ammonia industry can be estimated as follows. The ideal specific hydrogen demand for ammonia synthesis is 3 moles of H\textsubscript{2} for 2 moles of NH\textsubscript{3} \citep{oei_874}, or 177.55 kg\textsubscript{H2} per ton of ammonia.
        
        We acquire a list of all ammonia producers in Germany from the Industrial Association Agrar \citep{IVA.2018}. The production volumes of ammonia in Germany have been approximately constant since 2012 \citep{VCI.2020}. While, to the best of our knowledge, no information on site-specific current ammonia production is publicly available, we identify site-specific production capacities based on \cite{skwPiesteritz.2015, Peters.2016, BezirksregierungKoln.2017, Rechenberger.2020}. 
        The sum of these capacities (2,955,000 t/a) is somewhat higher than the current total ammonia production (2,415,327 in 2019). However, global ammonia demand is assumed to increase by 2030 \citep{hebling2019wasserstoff, IEA.2019}. Therefore, in the following, the production capacities are assumed as basis for the site-specific hydrogen demand estimation. 
        
        Regarding self-supply, no information on large-scale electrolysers at the identified ammonia plants was found. While \cite{BASF.2020} is building its own electrolysis plant for research purposes, no details are available about any large scale operation.  Correspondingly, the total demand is assumed to be equal to the net demand. With the assumptions made above, the site-specific demand can be estimated with equation \ref{equ:ammonia}. 
        \begin{equation}
        \begin{split}
            \label{equ:ammonia}
            HD = t_{Ammonia} * 177.55kg/t_{Ammonia} \\ * 33.33 kWh_{H_2}/kg_{H_2}
        \end{split}
        \end{equation}
            
    \paragraph{\textbf{Methanol}}
        The specific hydrogen demand is estimated as 2 moles of H\textsubscript{2} for 1 mole of CH\textsubscript{3}OH \citep{Hofbauer2016}, or 188.73 kg\textsubscript{H2} per ton of methanol. This is consistent with the assumptions of \cite{bazzanella2017low} and \cite{michalski2019wasserstoffstudie}.
        Currently, there are five relevant methanol plants in Germany \citep{Frohlich.2019}. However, one of them has terminated production and is being liquidated \citep{Thoma.2020}, and therefore is disregarded for 2030.
        
        In the next step, production capacities of the individual plants are identified \citep{fleiter2013energieverbrauch,Jendrischik.2020,BP.2019}. The sum of current production of 1,398,146 t/a \citep{VCI.2020} is lower than the total production capacity of 1,865,000 t/a. However, production has been rising in recent years, and global methanol demand is assumed to increase by 2030 \citep{hebling2019wasserstoff, IEA.2019}. Correspondingly, as with the hydrogen estimate for ammonia, the production capacity is used as basis for further calculations.
        
        Regarding self-supply, there are smaller electrolyzers for research purposes \citep{Jendrischik.2020, BASF.2020}, but no information on large-scale electrolysers at the identified ammonia plants was found. Correspondingly, the total demand is assumed to be equal to the net demand. With the assumptions made above, the site-specific demand can be estimated with equation \ref{equ:methanol}. 
        \begin{equation}
        \begin{split}
            \label{equ:methanol}
            HD = t_{Methanol} * 188.73kg/t_{Methanol} \\ * 33.33 kWh_{H_2}/kg_{H_2}
        \end{split}
        \end{equation}
            
    \paragraph{\textbf{Refineries}}
        We use the list of all refineries and their output capacities from the German Petroleum Industry Association \citep{MWV.2020}. Mineral oil consumption will decrease by varying degrees by 2030, depending on assumptions about the demand for liquid fuels \citep{michalski2019wasserstoffstudie}. Correspondingly, the current production volume of 87,013,000 tons is distributed across the sites in proportion to their processing capacity. Then, derived from the results of \cite{PrognosAG.2020b}, the assumption is made that the demand for mineral oil will decrease by about 20~\% until 2030. 
                    
        The specific hydrogen net demand is assumed to be approximately 100 $m^3_{H_2}$ per ton crude oil, based on \cite{schweer2002site}. Thus, the site-specific hydrogen demand for refineries can be estimated with equation \ref{equ:refineries}.
        \begin{equation}
        \begin{split}
        \label{equ:refineries}
            HD = t_{Oil,pq,2030} * 100 m^3/t_{Oil,pq,2030} \\ * 0,0841 * kWh_{H_2}/kg_{H_2} *22\%
        \end{split}
        \end{equation}

\section{Appendix B: Conversion factors} \label{appendix_B}

    \begin{table}[pos=htbp]
    \caption{Numeric values and conversion factors for H\textsubscript{2}} 
    \label{tab:hydrogen_conversion_factors}
        \begin{tabular}{ll}
        \toprule
        Lower heating value of hydrogen & 33.33 kWh/kg  \\
        Conversion factor kg in m³ & 11.89  \\
        Conversion factor m³ in kg & 0.0841 \\ 
        \bottomrule
        \end{tabular}
    \end{table}

\end{document}